\documentclass[12pt]{article}
\usepackage[utf8]{inputenc}
\usepackage{amsmath}
\usepackage[a4paper,
    left=30mm,
    right=30mm,
    top=20mm,
    bottom=20mm,
    ]{geometry}

\usepackage{titlesec}

\usepackage{lipsum}
\usepackage{booktabs}
\usepackage{url}

\usepackage{graphicx}
\title{Towards High Throughput Simulations and Machine Learning to Predict Polymer Phase Behaviour}
\author{Lois Smith}
\date{2024}

\usepackage{amssymb}
\usepackage{pifont}

\newcommand{\Lagr}{\mathcal{L}}
\newcommand{\un}{\mathcal{U}}
\newcommand{\ora}{\mathcal{O}}
\newcommand{\hyp}{\mathcal{H}}
\newcommand{\qu}{\mathcal{Q}}

\usepackage{graphicx}
\usepackage{float}
\usepackage{tabularx}
\usepackage{longtable}
\usepackage{array}
\usepackage{xltabular}
\usepackage{makecell}

\usepackage[doublespacing]{setspace}
\usepackage{cite}
\usepackage{chngcntr}
\counterwithin{table}{section}
\counterwithin{figure}{section}

\usepackage{subcaption}
\usepackage[doublespacing]{setspace}
\usepackage{cite}
\usepackage{amsmath}
\usepackage{enumitem}

\usepackage{graphicx}
\usepackage{float}
\usepackage{tabularx}
\usepackage{longtable}
\usepackage{array}
\usepackage{xltabular}
\usepackage{makecell}

\usepackage[colorlinks=false, urlcolor=blue,hidelinks]{hyperref}

\usepackage[utf8]{inputenc}
\usepackage{amsmath}
\usepackage[a4paper,
    left=30mm,
    right=30mm,
    top=20mm,
    bottom=20mm,
    ]{geometry}

\usepackage{chngcntr}
\counterwithin{table}{section}
\counterwithin{figure}{section}

\usepackage{caption} 

\usepackage[doublespacing]{setspace}
\usepackage{cite}

\begin{document}

\huge
 \begin{center}
Active Learning for Predicting Polymer/Plasticizer Phase Behaviour\\ 

\hspace{10pt}

\small
Lois Smith$^1$, Jessica Steele$^1$, Hossein Ali Karimi-Varzaneh$^2$, Paola Carbone$^1$ \\

\hspace{10pt}

\small  
$^1$ Department of Chemistry, School of Natural Sciences, The University of Manchester, M13 9PL, Manchester, United Kingdom \\
$^2$ Continental Reifen, Deutschland GmbH, D-30419 Hanover, Germany \\

\end{center}

\hspace{10pt}

\normalsize
\noindent Plasticisers are small additives commonly incorporated into polymer composites to enhance processability and improve mechanical properties. Their effectiveness depends heavily on their miscibility within the polymer melt, yet isolating the influence of plasticiser properties, such as flexibility and geometry, remains challenging. This difficulty stems not only from the time consuming nature of experimental work but also from the presence of impurities and inconsistencies that often arise during synthesis and testing. Atomistic simulations face similar difficulties as phase separation can occur over microsecond timescales, which can be computationally expensive. In this work, we use a coarse-grained bead-and-spring model to screen plasticisers of varying flexibilities and geometries to build a pool-based active learning procedure which characterizes their design space and its effect on polymer/plasticiser miscibility. We first sample a diverse unlabelled pool of data from a PL combinatorial space spanning millions of data points, and then perform an active learning cycle with a random forest model and an uncertainty/random hybrid query strategy to determine the miscibility behaviour (label) of queried molecules. This label is evaluated through computationally expensive, coarse-grained polymer/plasticiser simulations of a cis-(1,4)-polyisoprene melt filled with small hydrocarbon additives of varying sizes and rigidities. Through this, we are able to efficiently improve model performance in order to make predictions on the entire PL design space. We achieve a final F1 score of 0.89, indicating good model performance, and predict plasticiser immiscibility in two main regions of interest: rigid plasticisers with long backbones and bulky or frequent side chains, and plasticisers with flexible backbones but very rigid bulky side chains. Such findings enable us to determine a new set of general plasticiser design rules, suitable for non-polar molecules, which expands on our previous work. To further prove this, we perform atomistic simulations of polyisoprene/plasticiser systems which are approximately back-mapped from their coarse-grained equivalents. Our findings indicate that the polyisoprene/plasticiser phase behaviour, observed using the coarse-grained model for plasticisers in the absence of side chains, is valid.

\section{Introduction} 

\noindent In the past decades, polymers have emerged as a highly versatile and pervasive class of materials used for applications such as gas separation membranes~\cite{ROBESON1999549,Lasseuguette2022-zy,SIDHIKKUKANDATHVALAPPIL2021103}, energy storage devices~\cite{Oyaizu2024,LI2023101714,C4RA15947K}, packaging~\cite{TAJEDDIN2020525} and automotive components.~\cite{Patil2017,YUE2022107584} For such applications, there are a wide range of properties which must be carefully tuned, including thermal and electrical conductivity, selectivity, permittivity, corrosion resistance and glass transition temperature, $T_g$~\cite{YUE2022107584,doi:10.1021/ma9814548,Deshpande2014,li2023ionic,parin2024durability}; all of which impact the performance and processability of the material. 
As these requirements grow increasingly complex and the demand for better optimised polymers for target properties becomes greater, there has been a paradigm shift from traditional trial-and-error approaches towards informatics-driven prediction and design~\cite{Li,CHEN2021100595,audus,Struble2024}. Since polymer chemical and morphological spaces are incredibly vast, this allows for the targeted searching and identification of polymers most likely to fit design requirements. In such workflows, researchers attempt to establish robust structure-property relationships for the fast screening of large candidate search spaces. To this end, using curated datasets, they construct machine learning (ML) models capable of learning complex, non-linear relationships between a range of polymer chemical and morphological features, and properties of interest. Such methodologies have seen recent success in predicting the $T_g$s,~\cite{tao2021machine,park2022prediction,genome,casanola2024machine} thermal conductivities,~\cite{wu2019machine,huang2023exploring} dielectric constants,~\cite{chen2020frequency,genome,liang2021machine} bandgaps~\cite{genome,xu2021machine,alzahrani2024machine} and refractive indices~\cite{genome,lightstone2020refractive} of a wide range of polymers, among other relevant properties.~\cite{park2022prediction,sivaraman2020machine,genome,Jiang2024,ethier2023integrating} \par
\noindent There are many challenges associated with the machine learning approach. From the start, the representations of the morphology and chemistry of the polymer used as inputs in the model must be carefully selected. Polymers display features over a large range of length scales and are heavily influenced by additional complexities such as molecular weight distribution, the presence of additives or impurities and processing conditions.~\cite{gentekos2019controlling,yang2023identification,chandran2019processing,AMBROGI201787} Using polymer features that have poor correlation to the target feature can result in worse-performing models and so domain expertise is highly valued to guide initial data curation and feature selection. As a somewhat obvious example, the polymer radius of gyration, $R_g$, is strongly linked to degree of polymerization through Flory’s scaling law: $R_g \sim N^{\nu}$, where $\nu$ is the Flory exponent and, indeed, such a descriptor has emerged as very relevant in ML models used to predict $R_g$.~\cite{RN78} \par
\noindent Since the quality of a ML model is heavily dependent on the quality and availability of data, this presents a further obstacle to research. Despite recent efforts to curate large databases of polymer data, such as PolyInfo,~\cite{polyinfo} which has around 100 properties documented for over 33,000 polymer materials, including composites and blends, or to extract data from literature using natural language processing,~\cite{SHETTY2021101922} it still remains a challenge to obtain large amounts of uniformly formatted data which contain all descriptors of relevance for the target property. High-throughput (HTP) experiments offer a solution to this problem, as they can be tailored specifically to the polymer properties of interest and, in the case of HTP computation, the data can be automatically created and stored in a machine readable format with consistent polymer representations.~\cite{luo2024machine,marti,MA2022100850,nanjo2025spacier,Jiang2024} Computational techniques, such as molecular dynamics (MD), also offer a flexible framework for screening \textit{generic} polymer properties via coarse-graining, which can then be combined with experimental procedures to validate their findings. For instance, in 2024 Jiang and co-workers created a generative ML model to predict the $R_g$ of various linear, cyclic, star, branched, comb and dendrimer polymer topologies.~\cite{Jiang2024} They used a coarse-grained bead-and-spring model to quickly simulate 1342 generic polymers representing their chosen topological classes, and trained a variational autoencoder to compress their graph representations into a lower dimensional latent space. A regression auxiliary task was performed on the space to predict $R_g$ and a similar classification task to predict topological classes. This allowed them to create a physically meaningful latent space from which they could sample to create new polymer topologies with target $R_g$ values. They then performed a rheological analysis of the new polymers which found a link between their topology and viscosity behaviour. Such links are difficult to establish due to the entanglement of other factors affecting the rheology,~\cite{martini2018review,van2021role} highlighting a key advantage of their generic, coarse-grained approach. \par
\noindent The work of Jiang et al. performed the data labelling for their supervised learning on systems of single polymer chains (to characterize $R_g$), which has low computational cost. However, there are a wide range of polymer properties currently of practical interest which are not so simple to calculate.~\cite{adeyemi2022equilibrium,critical,roadmap} An additional dimension of complexity is added for polymer composites, which are polymers mixed with small additive particles or larger filler particles. Where the pure polymer may not satisfy the full property requirements for the application of interest, the additives offer a means to tailor its characteristics to closely match the target specifications. In the tyre industry, this is typically achieved by adding carbon black or silica filler particles which reinforce the polymer matrix, increasing its mechanical strength and durability.~\cite{RN52,RN53,Edwards1990,RN51,RN56,RN47,wolff1992filler} In addition to this, small petroleum-based plasticiser (PL) additives are added which sit in between polymer chains and increase their free volume, enhancing their mobility and, in turn, improving material flexibility and processability.~\cite{RN111,RN115,RN57} Concern over environmental impact and sustainability has spurred on the discovery and testing of new PLs in recent years.~\cite{RN99, RN101,RN112, RN113,RN114} \par

\noindent The miscibility of such molecules affects their compatibility with the polymer and is of paramount importance to their success as additives. Thus, a large amount of work has been performed to investigate and characterise such a property in industry relevant polymers,~\cite{RN72,RN71,RN69,RN121} however this is a time-consuming task. In our previous work, we established a route to the HTP screening of generic hydrocarbon PLs with different topologies, varying their degree of flexibility, side chain lengths and degree of branching, and linking such properties to their miscibility in a chemically specific coarse grained cis-(1,4)-polyisoprene (PI) melt~\cite{smith2024framework}. We demonstrated that a decision tree model was capable of creating a design path towards miscible PL topologies, using an array of simple-to-calculate geometric and thermodynamic PL descriptors, even with a small design space, thus paving the way for a more complex learning procedure, varying a larger amount of PL features. \par
\noindent In this latest work, continuing this generic approach to PL design, we develop a pool-based active learning (AL) procedure with a predictive random forest (RF) model to characterize PL miscibility throughout an expanded search space. Such a pool-based approach, along with other AL algorithms, have been used previously to uncover polymers with enhanced thermal conductivity,~\cite{Xu2024,zhang2024active} polymer dielectrics with high bandgaps~\cite{KIM2021110067,mannodi2016machine} and polymers with a specified range of $T_g$.~\cite{Jiang2024} Their main draw is in applications where data labeling (measuring the target property) is expensive, which severely limits data collection ability, as they are able to smartly search feature spaces for data points which benefit a surrogate ML model the most via a query strategy.~\cite{smith2018less,zhang2019bayesian} \par
\noindent Considering this, we greatly expand the PL search space from our previous work to include a wider variation of PL flexibilities, as well as differing the relative flexibility of the PL backbone and its side chains, which we envision represents molecules of differing degrees of unsaturation. We further widen the range of PL backbones, which was not varied in our previous work, and allow full variation of side chain lengths within a single PL molecule.
We additionally provide a validation for our coarse grained representation by measuring the phase behaviour of hydrocarbon PLs in an atomistic PI melt, with topological features approximate to their coarse grained counterparts.  \par

\section{Methodology}

\noindent \textit{\large{Coarse-grained to Atomistic PL Mapping}} \par

\noindent The coarse grained PLs used in this and previous works~\cite{RN84,RN116,smith2024framework} represent generic hydrocarbons with varying flexibilities and geometries. As such, there is no direct mapping to their atomistic counterparts. Therefore, to validate the phase behaviour of the coarse grained model, we select hydrocarbons which display a general similarity to coarse grained PLs based on the properties which affect their tendency to phase separate, and study their behaviour within an atomistic PI melt. \par

\begin{figure}[H] 
    \centering
    \includegraphics[width=0.85\textwidth]{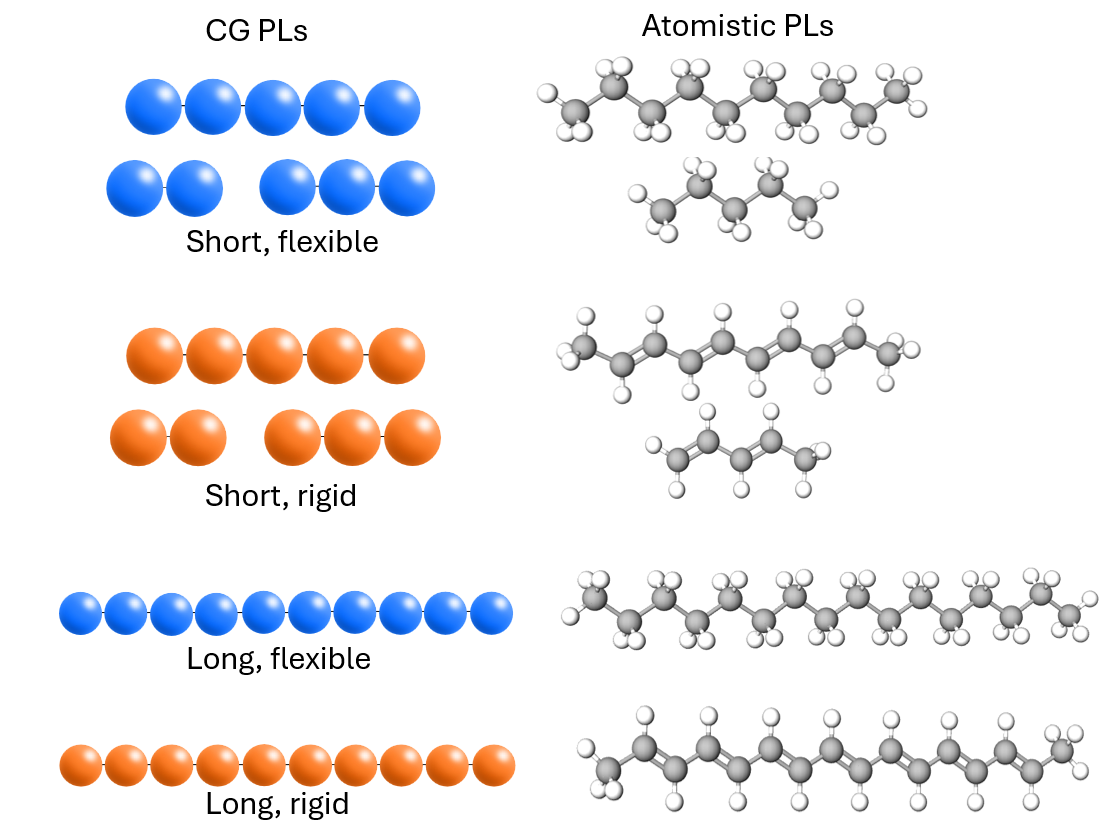} 
    \captionsetup{font={stretch=1.5}}
    \caption{Diagram displaying an approximate mapping between the coarse grained PL representation and atomistic counterparts. Fully flexible PLs are depicted in blue, rod-like PLs are depicted in orange. Corresponding atomistic molecules (top to bottom) are decane, pentane, 1,3,5,7,9-decapentaene, 1,3-pentadiene, hexadecane and 1,3,5,7,9,11,13,15-hexadecoctaene.}
    \label{fig:mapping} 
\end{figure}

\noindent The most simplistic molecules studied in our initial work on PL phase behaviour~\cite{RN116} are linear and either rod-like or flexible. Thus, to mimic short and long molecules, we use short and long chain alkanes. To further differentiate between those with a high and low degree of flexibility, we vary the degree of saturation; where very rigid molecules have alternating double bonds (polyenes) and those that have a high flexibility are completely absent of them (alkanes). Figure~\ref{fig:mapping} shows such an approximate mapping from coarse grained to atomistic PLs. \par

\noindent The mapping was performed on the basis of molecular weight. Each coarse grained bead is 67$\%$ of the mass of a PI monomer, 45.64 $\text{g/mol}$ and, in our previous work~\cite{RN116}, we determined a molecular weight of 7 beads, 319.48 $\text{g/mol}$ was required to observe the cross-over from miscible to immiscible behaviour for rigid, linear PLs. Thus, mimic shorter PLs, we chose hydrocarbon molecules with a molecular weight much less than this value. Specifically, decane, pentane, 1,3,5,7,9-decapentaene and 1,3-pentadiene with molecular weights 142.286, 72.151, 134.222 and 68.119 $\text{kJ/mol}$ respectively. Long PLs are represented by the 16 carbon alkane, hexadecane, and polyene, 1,3,5,7,9,11,13,15-hexadecoctaene which have molecular weights of 226.448 and 212.336 $\text{kJ/mol}$. These are lower than the threshold value for the coarse grained model, but were chosen due to the large anticipated computational cost associated determining the PI/PL phase behaviour with simulating large molecules that diffuse slowly through the melt. \\ 

\noindent \textit{\large{Atomistic Model Details}} \par

\noindent To perform the atomistic simulations of cis-(1,4)-polyisoprene and hydrocarbon PLs, we use the OPLS/2020 model,~\cite{jorgensen2023opls,ghahremanpour2022refinement} which is an update on the OPLS/AA model to include a refinement for liquid alkanes. The new parameters improve on the heats of vaporisation of long-chain alkanes compared to the previous, and their transport properties confirm their correct liquid phase behaviour. OPLS/AA itself has previously been used for the simulation of cis-(1,4)-polyisoprene melts to study their mechanical and thermal properties, and can accurately reproduce the experimental $T_g$, heat of vaporisation and Hildebrand solubility parameter, $\delta$.~\cite{costa2024temperature,yolong2019local,chen2022molecular}
The non-bonded interactions are handled by the 12-6 Lennard-Jones potential, given by

\begin{equation}
  U(r)=\begin{cases}
    4\epsilon \left [\left(\frac{\sigma}{r}\right)^{12} - \left(\frac{\sigma}{r}\right)^6 \right], & \text{if $r<r_c$}.\\
    0, & \text{otherwise}
  \end{cases}
\end{equation}

\noindent where $\sigma$ is the energy well depth, $\epsilon$ is the distance at zero potential energy and $r_c$ is the cut-off distance. The bonded potential is 

\begin{equation}
    U_{bond}(r) = K_b (r-b_0)^2
\end{equation}

\noindent where $K_b$ is a stiffness constant and $b_0$ is the equilibrium bond distance. The angle and dihedral potentials are

\begin{equation}
    U_{angle}(\theta_{ijk})=K^{\theta}_{ijk}(\theta_{ijk}-\theta^0_{ijk})^2
\end{equation}

\begin{multline}
    U_{dihedral}(\phi_{ijk})=\frac{1}{2}K_1[1+\cos(\phi_{ijk})] + \frac{1}{2}K_2[1-\cos(2\phi_{ijk})] \\ + \frac{1}{2}K_3[1+\cos(3\phi_{ijk})] + \frac{1}{2}K_4[1-\cos(4\phi_{ijk})]
\end{multline}

\noindent respectively, where $K^{\theta}_{ijk}$, $K_1$, $K_2$, $K_3$ and $K_4$ are stiffness constants, $\theta_{ijk}$ and $\theta^0_{ijk}$ are the angles and equilibrium angles between atom triplets $i-j-k$ respectively, and $\phi_{ijk}$ is the angle between the $ijk$ and $jkl$ planes. A complete list of these parameters is provided by the OPLS/2020 forcefield.~\cite{jorgensen2023opls,ghahremanpour2022refinement} The MD simulation package GROMACS 2023~\cite{RN85,van2005gromacs} was used for all atomistic simulations in this work. \\

\noindent \textit{\large{Phase Behaviour of PI/PL systems}} \par

\noindent Determining the phase behaviour of atomistic polymer mixtures is a computationally costly process, and phase separation can occur on $\mu$s timescales. In our case, this is exasperated by the high viscosity of the PI and increasingly slow expected diffusion of PL molecules through its matrix with increasing PL molecular weight. Thus, to study the PL/PI phase behaviour by encouraging favourable PL aggregation, we inserted 80 32mer PI chains into a 12nm $\times$ 12nm $\times$ 12nm box, along with a pre-made aggregate of randomly arranged PLs at a concentration of 5 phr. The system was firstly minimised using the steepest-descents algorithm until an energy minimum was reached. Then, it was simulated for 10 ns in an NPT ensemble until a plateau in system density was observed. The temperature was kept constant at 298.1 K using the v-rescale thermostat with a time constant, $\tau_T$, of 1 ps; and the pressure was held at 1 bar using the c-rescale barostat with $\tau_p=5$ ps. The non-bonded interactions were truncated at 1 nm using the Verlet cut-off scheme, and smoothly switched to zero at 1.2 nm. The electrostatics were handled by the Particle-mesh Ewald (PME) scheme with a cut-off radius of 1.2 nm and a PME order of 6. Constraints were applied to all hydrogen bonds using the Lincs algorithm and the time step was set to 0.001 ps. Finally, periodic boundary conditions were applied in all dimensions to form an infinite box. The resulting output configuration from the NPT equilibration was then used for an NVT production run at 298.1 K for a total of 800 ns, or until PL equilibrium phase behaviour was observed via the plateau of the miscibility parameter, $\zeta$, used in our previous works~\cite{RN116,smith2024framework}, over 100 ns. \par

\noindent As an additional method of verification, we calculated the Hildebrand solubility parameters~\cite{barton1975solubility}, $\delta$, of PI, along with each PL, which is a common method used to predict the solubility of two component systems.~\cite{critical} $\delta$ predicts the solubility of molecules based on a ‘like dissolves like’ principal in which more similar chemistries are more likely to be compatible. It is defined by the square root of the cohesive energy density (CED) at room temperature, which is, qualitatively, the energy required to vaporize one mole of liquid substance to vacuum,

\begin{equation}
    \delta=\sqrt{\text{CED}} = \sqrt{\frac{\Delta U_{vap}}{V_M}}
\end{equation}

\noindent where $\Delta U_{vap}$ is the energy of vaporization of the substance and $V_M$ is its molar volume. \par
\noindent $\Delta U_{vap}$ can then be calculated by taking the difference of the potential energy of a molecule in bulk and a single molecule in vacuum:

\begin{equation}
    \Delta U_{vap} = E_{vac}-E_{bulk}
\end{equation}

\noindent Those species with $|\Delta \delta| < 2~\text{MPa}^{1/2}$ are typically considered good solvents. \par

\noindent Determining $E_{bulk}$ for the PL molecules was achieved by inserting 1200 molecules (in the case of pentane and 1,3-pentadiene) or 512 molecules (for the remaining PLs) into a 12nm x 12nm x 12nm simulation box and performing a 2ns NVT equilibration, followed by a 2ns NPT equilibration with the Berendsen barostat and a 10ns production run with the c-rescale barostat. To simulate the molecules in vacuum, 10 single molecules taken from the final configuration of the bulk simulation were placed in their own simulation boxes and simulated for 5 ns. The potential energy was averaged over all configurations. \par
\noindent To calculate the Hildebrand solubility parameter of the PI, we followed a similar procedure to that described by Costa and co-workers.~\cite{costa2024temperature}  \par

\noindent First, we randomly inserted 20 100mer chains into a simulation box of size 12nm x 12nm x 12nm and minimized to obtain a maximum force $<100~\text{kJ/mol}$ using the steepest descent algorithm. We then simulated in an NVT ensemble for 2 ns at 500~K. Subsequently, we simulated in an NPT ensemble at 500~K for 100~ns, and then annealed to 298.1 K at a rate of approximately 2 $\text{K}~\text{ns}^{-1}$, where it was simulated for a further 100 ns. From this, we obtained a density of 900.5 $\pm$ 0.21 $\text{kg/m}^3$, which is a slight underestimation of the experimental density of 910 $\text{kg/m}^3$~\cite{RN86} but is consistent with other computational works using cis-PI.~\cite{RN86,yolong2019local} \par

\noindent To simulate the PI in vacuum, we selected 10 chains from the last configuration of the melt and simulated each in vacuum for 2 ns. The coordinates of each chain were `frozen' to mitigate a known problem simulating large molecules in vacuum where they form coils due to self-interaction, misrepresenting their amorphous state structure. After each simulation, the potential energy was averaged over all chains. \\

\noindent \textit{\large{Coarse-Grained PLs}} \par

\noindent The coarse-grained PI and PLs simulated in this work uses the same parameterized Kremer-Grest bead-and-spring model as in our previous work,~\cite{svaneborg2016kremer,RN116,smith2024framework}. To impart additional degrees of flexibility to the PL molecules, we tuned the bending parameter, $k_h$, of the harmonic potential

\begin{equation} \label{eq:harmonic}
    U_{bend}(\theta) = k_h(\theta-\theta_0)^2
\end{equation}

\noindent where $\theta_0$ is the equilibrium angle between bead triplets and $k_h$ is the bending constant. Previously, a value of $k_h=100$ kJ/mol represented rod-like molecules, and their fully flexible counterparts used 

\begin{equation} \label{eq:kg-bend}
    U_{bend}(\theta) = \kappa(1-\cos \theta)
\end{equation}

\noindent where $\kappa = 0.6204~\text{kJ/mol}$. Now, we allow $k_h$ to vary between 100 and 20 in multiples of 10 to produce molecules with intermediate degrees of flexibility.\par

\noindent PL features such as backbone length, $L_{back}$, side chain length, $L_{side}$ and side chain grafting density, $\rho_{side}$, retain the same definitions as per our previous work~\cite{smith2024framework}, with the added restriction that side chains cannot be placed on the first and last beads of the backbone, nor on directly adjacent beads. The former was done to ensure no artificial backbone lengths occurred. Finally, the value of $L_{side}$ is restricted to a maximum length of $L_{back}-1$. \par

\begin{figure}[H] 
    \centering
    \captionsetup{font={stretch=1.5}}
    \includegraphics[width=0.9\textwidth]{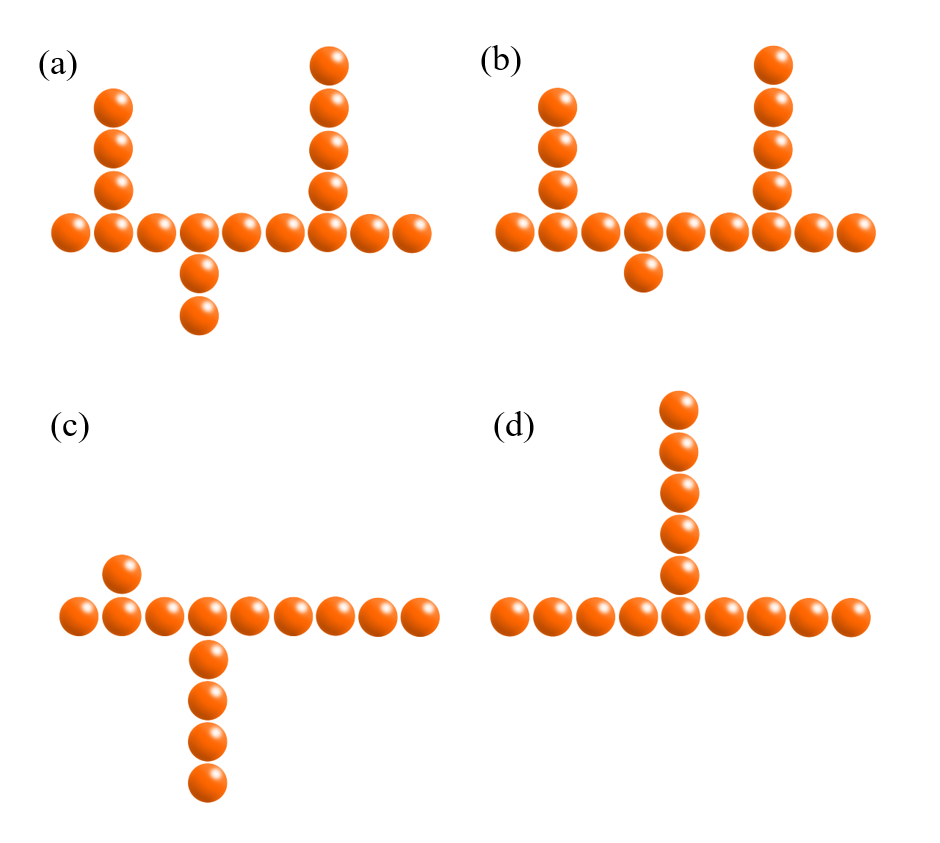} 
    \caption{Depictions of possible PL molecules allowed in this work. Side chains can be placed on any bead along the backbone, except for the first, last and in directly adjacent positions. Each has a maximum length of $L_{back}-1$.}
    \label{fig:pos-configs} 
\end{figure}

\noindent The backbone lengths were varied from 4 to 14 which we believe is sufficient to fully capture the crossover from miscible to immiscible behaviour, given its relationship to PL size, established previously.~\cite{smith2024framework} Side chains can then be placed in any allowed arrangement across the backbone, and can be any allowed length. Examples of possible PL molecules is displayed in Figure~\ref{fig:pos-configs}. \par

\noindent Each possible PL molecule can then be simulated at one of 10 degrees of flexibility, as determined by the value of $k_h$ in Equation~(\ref{eq:harmonic}). Additionally, we explore three distinct variations in intra-molecular flexibility, not considered in out previous work: (i) a rigid backbone with flexible side chains, (ii) a flexible backbone with rigid side chains, and (iii) a rigid backbone with alternating rigid and flexible side chains. Here the rigidity is imparted using Equation~\ref{eq:harmonic} 
with $k_h=100~\text{kJ/mol}$ and the flexibility using Equation~\ref{eq:kg-bend}. Full details on the PLs selected for AL can be found in Section~\textit{Active Learning Procedure}. \par

\noindent To obtain the miscibility labels for each PL molecule, we follow a procedure similar to that described in our previous work~\cite{smith2024framework}, of an NPT equilibration followed by NVT production run. However, for increased computational efficiency, we began each simulation with PLs placed randomly within an aggregate, instead of evenly dispersed throughout the simulation box. We then simulated in an NVT ensemble for 3$\mu$s, which we found was sufficient time to observe either dissolution or rearrangement of PLs into more optimally packed aggregates. Contrastingly, beginning with the PLs dissolved in the melt and waiting for aggregation can take up to 10$\mu$s, which poses a significant barrier to high throughput simulations. We also reduced the system size from 72 PI chains of length 300 to 285 chains of length 50, keeping the concentration of PLs at 5 phr, to reduce simulation time. Extensive internal tests performed verified that these modifications did not impact the observed miscibility behaviour. To quantitatively analyse it, we used the miscibility parameter, $\zeta$, developed in our previous work.~\cite{RN116}.   \par
\noindent To ensure a process with minimal human intervention, we automatically assign those PLs who have $\zeta < 2.7$ as miscible (from our previous work) and those with $\zeta>5$ as immiscible. A small number of PLs ($<$ 10$\%$) had miscibility of $2.7 < \zeta<5$ and manual analysis of their trajectories showed repeated formation and deformation of aggregates, akin to the meta-stable behaviour found in our previous work. Such simulations indicate either a deficiency in simulation time, or the existence of a third 'meta-stable' class. To rule out the former, we extended the systems by 3$\mu$s and indeed this provided definitive behaviour in the majority of cases. The few that displayed the same behaviour were discarded, as we anticipated they would not impact the generic design rules we investigate in this work, and our PLs also do not map directly onto atomistic counterparts which obscures the practicality of studying the meta-stable PLs in more detail. Thus, the task becomes a binary classification, for which we assign miscible = 0 and immiscible = 1. \par
\noindent We use the MD simulation package GROMACS 2018~\cite{van2005gromacs} for all coarse grained simulations in this work, and the HTP simulation manager, Signac,~\cite{signac_commat,signac_scipy_2018,signac_scipy_2021} with python 3.11.~\cite{10.5555/1593511} \\

\noindent \textit{\large{Active Learning Procedure}} \par

\begin{figure}[H] 
    \centering
    \includegraphics[width=0.9\textwidth]
    {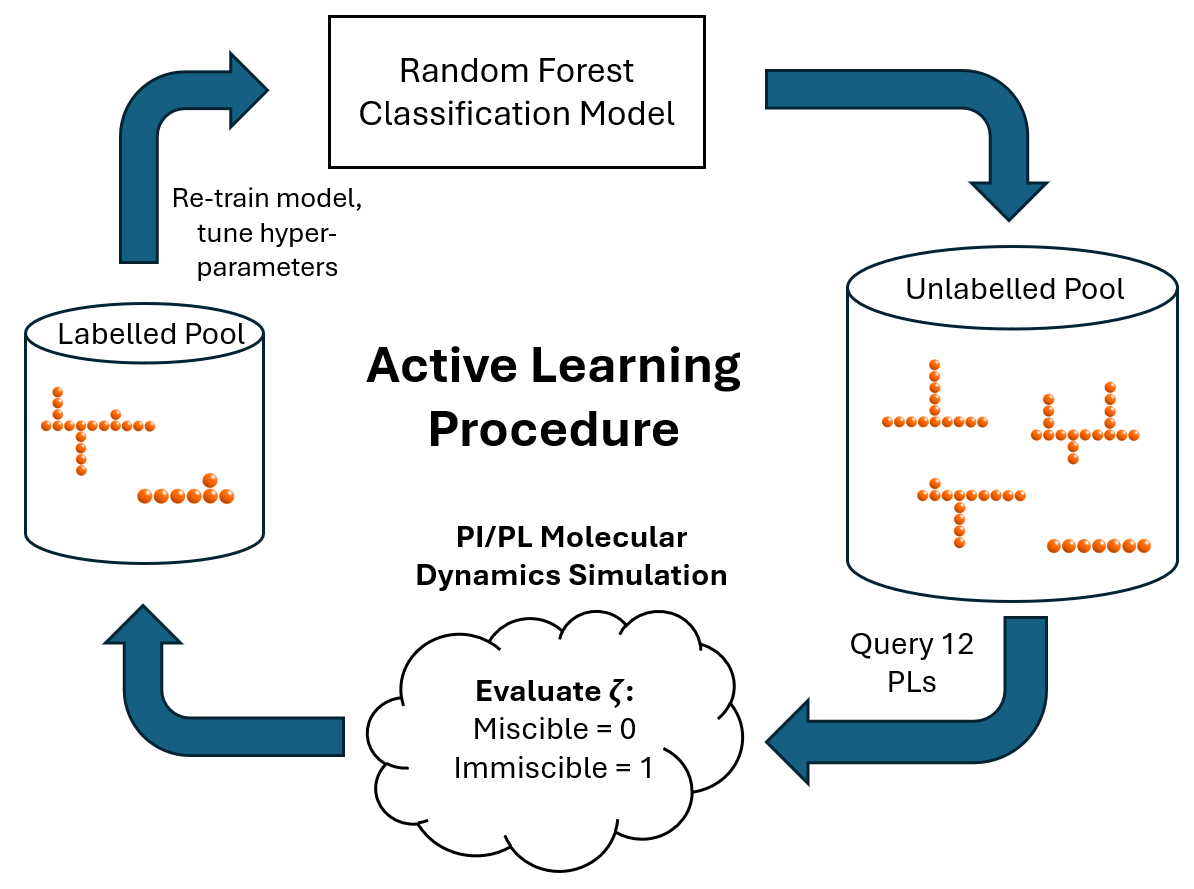} 
    \captionsetup{font={stretch=1.5}}
    \caption{Depiction of the pool-based AL approach used in this work. A RF classification model is iteratively updated by querying PL instances from a large, unlabelled pool of data. Data is obtained via a computationally expensive oracle, $\ora$, which is a $3~\mu$s PI/PL coarse grained simulation.}
    \label{fig:alearn} 
\end{figure}

\noindent The pool-based AL protocol assumes a large, unlabelled pool, $\un$, a smaller, labelled pool, $\Lagr$, a learner, and an oracle, $\ora$, which provides ground truth labels. Firstly, the learner is built on a small hypothesis set of data, $\hyp$, to estimate an expensive objective function, $f$, mapping $x_n\to y_n$. Then, each round of the procedure follows the steps,~\cite{Lu2023ReBenchmarkingPA} also depicted in Figure~\ref{fig:alearn}:

\begin{enumerate}[leftmargin=.4in]
\item Use a query strategy $\qu$ to select an instance, $x_n$, from $\un$. Since labeling our data is computationally expensive, we chose to batch query with a batch size of 12. 
\item Use $\ora$ to label the data, $\ora(x_n)=y_n=\zeta$. This is a PI/PL simulation as described in Section~\textit{Coarse-Grained PLs}.
\item Update the labelled pool to include the new data: $\Lagr \gets \Lagr ~\cup \{(x_n,y_n)\} $, $\un \gets \un ~ $\textbackslash$ ~\{(x_n)\}$.
\item Update the learner by training on the new labelled pool, $\Lagr$.
\end{enumerate}

\noindent The initial unlabelled pool, formed by the combinatorial space spanned according to the PL selection criteria outlined in Section~\textit{Coarse-Grained PLs} is large ($>$10,000,000) and sparse. For example, PL molecules with small backbones are underrepresented, compared to those with large backbones, because of the fewer number of allowed side chain placements. Additionally, many PLs have a high degree of similarity to one another, such as those with a side chain differing in length or placement of one of two beads, an example of which can be seen in Figures~\ref{fig:pos-configs}(a) and (b). Since this work is focused on the generation of generic PL design rules, we do not require an exhaustive sampling of highly similar PLs, and their inclusion would not significantly enhance the diversity of the design space. Thus, we chose to selectively sample PLs from the total design space that display diversity over key molecular features: backbone length, $L_{back}$, side chain number, $N_{side}$, average side chain length, $<L_{side}>$, backbone flexibility, $k_{back}$, (represented through the value of the bending parameter, $k_h$), and average side chain flexibility, $<k_{side}>$. \par

\noindent Such a reduced design space offers the advantage of lower computational cost associated with its analysis, without compromising the ability to extract meaningful and generalisable PL design principles. Thus, we selected 2055 PLs to form the unlabelled data pool. Further details on the design space sampling for the unlabelled pool can be found in the Supporting Information.   \par

\noindent Our previous work identified two geometric descriptors, the square radius of gyration, $R_g^2$, and acylindricity, $c$, which correlate with PL miscibility. Our findings indicated that larger PL molecules had a tendency to agglomerate compared to their smaller counterparts. Their size is directly linked to their spatial features, such as $L_{back}$, $<L_{side}>$, and degree of flexibility, since more flexible molecules can adopt more compact conformations within the PI melt. With this in mind, we chose to describe each PL molecule by its $L_{back}$, $<L_{side}>$, side chain number, $N_{side}$, backbone flexibility, $k_{back}$ and average side chain flexibility, $<k_{side}>$; as such features almost fully describe effective PL size and do not require a simulation to calculate. \par 

\noindent The initial learner was then built from 116 randomly selected data points, 47 of which were used for the initial train set, and 57 for the hold-out test set, which is used to assess the performance of a RF surrogate model at each stage of the AL, via the model accuracy and F1 score. Also at each stage, the hyperparameters: number of trees, maximum tree depth, minimum data points per node split and minimum data points per leaf node, were also tuned via grid search with 4-fold stratified cross-validation. Since the initial amount of labelled data is small, there is a high risk of overfitting. To mitigate this, we compared the F1 score of model predictions on both the training and validation data. Models displaying a large discrepancy ($>$0.1) were discarded in favor of those with hyperparameters that resulted in more consistent performance. \par

\noindent After each round of active learning, we use a hybrid query strategy to select new points for labelling. Of the 12 points queried, 7 are chosen through uncertainty sampling. To do so, we evaluate the model's entropy of the unlabelled pool:

\begin{equation} S =-\sum_i p_i \ln p_i \end{equation}

\noindent where $p_i$ represents the probability of the point belonging to either the miscible or immiscible class. These probabilities are calculated based on the vote fractions of the trees in the random forest. By prioritizing points with the highest entropy, we target regions where the model is most uncertain, helping to refine decision boundaries and improve performance. \par

\noindent The remaining 5 points are selected randomly from the unlabelled pool. This introduces diversity into the training set and encourages exploration of underrepresented regions in the search space. By balancing exploitation (uncertainty sampling) with exploration (random sampling), this strategy enhances the model’s ability to generalise effectively. We choose to terminate the procedure if we achieve a consistent F1 score over 3 rounds of active learning. Thus, if the data acquisition ceases to meaningfully benefit the surrogate model, we avoid redundant simulations.

\section{Results}

\noindent \textit{\large{Atomistic PL Miscibility}} \par

\noindent To assess the miscibility of the PI/PL systems, we used the average miscibility parameter, $\zeta$, calculated with the PL-PL centre of mass radial distribution function, extracted over a region where PI/PL phase equilibration had been reached. Table~\ref{tab:zeta} displays the results.

\begin{table}[H]
\centering
\captionsetup{font={stretch=1.5}}
\caption{Table displaying the value of the miscibility parameter, $\zeta$, of the atomistic PLs simulated in this work mixed in a PI matrix.}
\label{tab:zeta}
\begin{tabular}{ccc} \toprule
    {PL name} &
    {SMILES} &
    \makecell{$\zeta$} \\
    \midrule
    pentane  & \texttt{CCCCC} & $1.06~\pm~0.003$ \\
    dodecane & \texttt{CCCCCCCCCCCC} & $1.21~\pm~0.01$ \\
    hexadecane & \texttt{CCCCCCCCCCCCCCCC} & $1.77~\pm~0.01$ \\
    1,3-pentadiene & \texttt{C=CC=CC} & $1.06~\pm~0.003$  \\
    1,3,5,7,9-decapentaene & \texttt{C=CC=CC=CC=CC=C} & $8.87~\pm~0.03$ \\
    1,3,5,7,9,11,13,15-hexadecoctaene & \texttt{C=CC=CC=CC=CC=CC=CC=CC=C} & $13.15~\pm~0.01$ \\
    \bottomrule
\end{tabular}
\end{table}

\noindent The results indicate that completely saturated hydrocarbon molecules of increasing molecular weight (alkanes) are miscible ($\zeta \approx 1$) within the PI melt. The addition of a high degree of double bonds in the structure does not impact the miscibility of the molecule (1,3-pentadiene, 1,3,5,7,9-decapentaene), except for those of high molecular weight (1,3,5,7,9-decapentaene, 1,3,5,7,9,11,13,15-hexadecoctaene) which are immiscible within the melt. This supports our findings of the previous work, that PLs in the absence of side chains are miscible regardless of molecular weight when flexible, and immiscible when rigid and of a high molecular weight.~\cite{RN116} \par 

\noindent Table~\ref{tab:hild} displays the Hildebrand parameters for both the PI and PL molecules. Comparing the calculated Hildebrand parameters with available experimental values, a consistent underestimation exists. Particularly, PI has an experimental value of $\delta$ of 16.3-17.8~$\text{MPa}^{1/2}$, compared to $\delta =14.6~\text{MPa}^{1/2}$ from our results. This underestimation to a slightly lesser degree is also present in pentane, dodecane and hexadecane. Our results also show underestimation compared to another work calculating $\delta$ for cis-PI and hydrocarbon solvents using the OPLS/AA forcefield. Despite this, the solubility predictions of our results, shown by the value of $\delta_{PI}-\delta_{PL}$ in Table~\ref{tab:hild}, fall in line with those provided by the available experimental data. They also agree with the all of the PI/PL simulations we performed, as shown by the $\zeta$ values reported in Table~\ref{tab:zeta}. In Table~\ref{tab:hild}, we also report $\delta$ for 3 additional branched alkanes and polyenes with flexible side chains of varying lengths. The prediction of such regions of the PL design space indicate the addition of such flexible side chains should hinder aggregation. Indeed, this is in line with the prediction indicated by the difference in $\delta$ between PI and the PLs. Thus, we have validated the behaviour of both linear coarse-grained PLs of varying molecular weight and rigidity, and those with flexible branches, with their approximate atomistic counterparts, illustrating the applicability of our work to atomistically detailed systems. \\

\begin{table}[h!]
\centering
\captionsetup{font={stretch=1.5}}
\caption{Table showing the $\delta$ values of atomistic PI and hydrocarbon PLs, along with their differences ($\Delta \delta$) and available experimental values ($\delta_{exp}$). The errors in PL $\delta$ values are negligible, as they are an order of magnitude smaller than the reported precision.}
\label{tab:hild}
\begin{tabular}{cccc} \toprule
    {PL/polymer name} &
    \makecell{$\delta$ \\ (MPa$^{1/2}$)} & 
    \makecell{$\Delta \delta=\delta_{PI}-\delta_{PL}$ \\ (MPa$^{1/2}$)} &
    \makecell{$\delta_{exp}$ \\ (MPa$^{1/2}$)} \\
    \midrule
    cis-(1,4)-polyisoprene & $14.6~\pm~0.2$ & - & 16.3-17.8~\cite{barton2018handbook} \\
    pentane  & $13.5$ & $~1.1$ & 14.5~\cite{imre2003effect} \\
    dodecane  & $15.1$ & $-0.5$ & 16.0~\cite{imre2003effect}\\
    hexadecane  & $15.0$ & $-0.4$ & 15.9~\cite{yaws2008thermophysical,lide2004crc}\\
    1,3-pentadiene & $15.4$ & $-0.8$ & N/A\\
    1,3,5,7,9-decapentaene & $23.1$ & $-8.5$ & N/A \\
    1,3,5,7,9,11,13,15-hexadecoctaene  & $24.1$ & $-9.5$ & N/A\\
    3,6,9,12-tetrahexyl-15-\\methylheneicosane & $16.4$ & $-1.8$ & N/A \\
    3,6,9,12-tetrabutyl-15-methyl-\\1,3,5,7,9,11,13,15-nonadecaoctaene & $16.9$ & $-2.3$ & N/A \\
    3,6,9,12-tetraoctyl-15-methyl-\\1,3,5,7,9,11,13,15-tricosaoctaene & $16.4$ & $-1.8$ & N/A \\
    \bottomrule
\end{tabular}
\end{table}

\noindent \textit{\large{Active Learning}} \par

\begin{figure}[H] 
    \centering
    \includegraphics[width=0.9\textwidth]
    {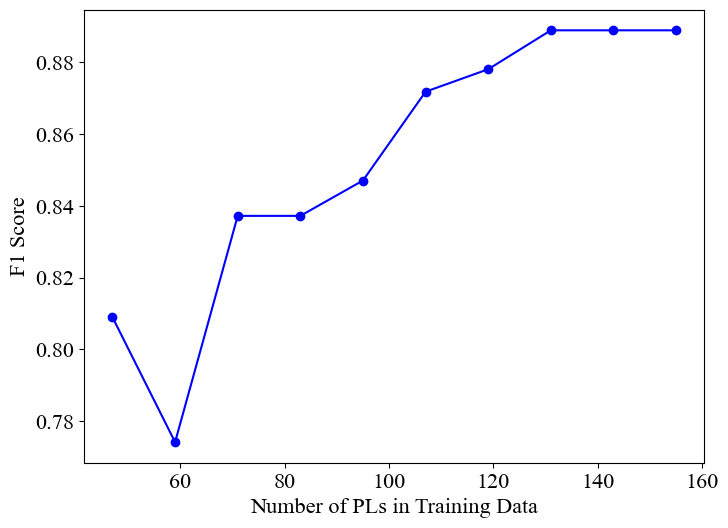} 
    \captionsetup{font={stretch=1.5}}
    \caption{Plot of the random forest classifier F1 score as a function of the number of PL data points included within the training the data. The peak F1 score reached is 0.89.}
    \label{fig:al-perf} 
\end{figure}

\noindent The RF surrogate model F1 score as a function of training data size is provided in Figure~\ref{fig:al-perf}. From this, we observe a high performance with the hold-out test set upon the inclusion of 155 data points, with an F1 score of 0.89. To contextualise this performance, we constructed a dummy model from the training data of the final round of AL, which predicts the label of the majority class (miscible) in the data. Such a model provides a baseline performance metric which our model should exceed in order to make useful predictions. The dummy model has an F1 score of 0.77 and so is significantly worse performing compared to our model. \\

\begin{figure}[H] 
    \centering
    \includegraphics[width=0.95\textwidth]
    {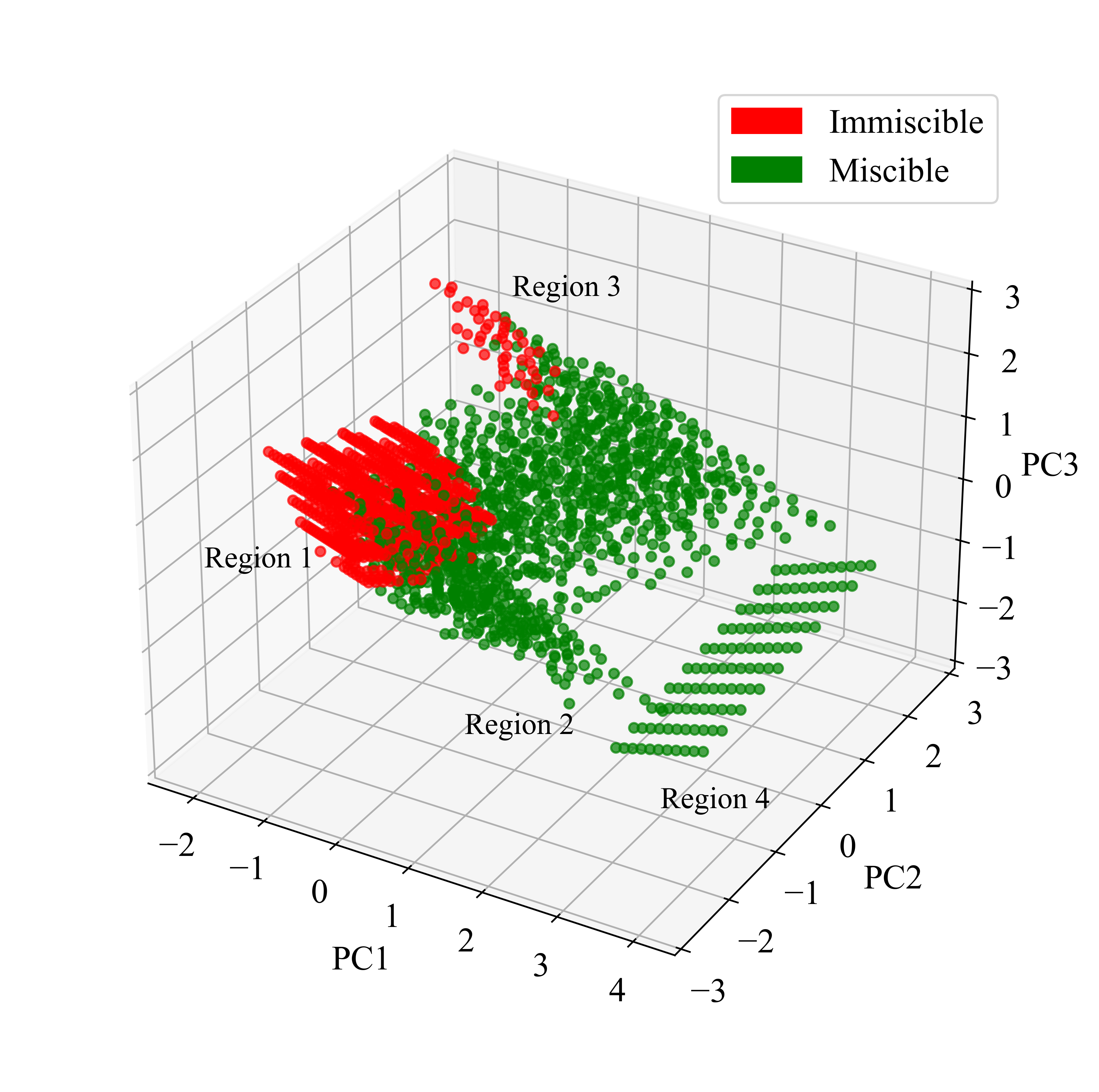} 
    \captionsetup{font={stretch=1.5}}
    \caption{Full PL data space expressed in 3 principal components. The points coloured in green and red represent PLs which the RF model predicts are miscible and immisicble, respectively.}
    \label{fig:predict} 
\end{figure}

\noindent Using the final model of the AL procedure, we can make predictions on the entire design space. Figure~\ref{fig:predict} shows the data space projected onto a lower dimensional space, using pricipal component analysis (PCA), coloured by the predicted miscibility behaviour. From this, we can firstly observe two main areas where PLs display predicted immiscibility (Region 1 and Region 3). The first, and largest, area primarily represents PLs of high backbone and side chain rigidities ($k_{back}$, $k_{side}$ = 70-100). As both rigidities decrease, we observe a cross-over from immiscible to miscible PL behaviour. This region also consists of typically larger PLs, with higher values of $L_{back}$ and $<L_{side}>$. This can be seen visually by observing the data space coloured by such features, as shown in the Supporting Information. This is expected given that, in our previous work, we observed a strong correlation between PL size and rigidity with its miscibility behaviour. \par

\noindent A further area of interest (Region 2 in Figure~\ref{fig:predict}) occurs with PL molecules with a high degree of backbone rigidity and low degree of side chain rigidity, where complete miscibility is predicted. This can be rationalised by considering that the high flexibility of the side chains reduces the effective size of the molecules, as they allow it to access more compact conformations. Conversely, in the opposite case, the second region of predicted immiscibility (Region 3 in Figure~\ref{fig:predict}) consists of PLs with a low degree of rigidity in their backbone and a high degree of rigidity in their side chains. The side chains in this space also have a typically high values of $<L_{side}>$. Thus, we can infer that the presence of long, rigid side chains are an important factor in encouraging aggregate formation, regardless of the rigidity of the backbone. \par

\noindent In the final sub-type of PL, which are those with rigid backbones and alternating rigid and flexible side chains, we observe complete predicted miscibility from our model. This is likely due to the requirement of alternating side chain flexibility preventing a significant amount of PL molecules in the data possessing enough rigid side chains to induce PL aggregate disruption, since we have already deduced that the presence of long, flexible side chains do not induce cluster dissolution. Thus, the model may not be sufficient at predicting this area of the feature space. 

\noindent Important to discuss is a likely failure of our current model to predict the small, yet significant area of the feature space consisting of PLs with no side chains (Region 4). Such configurations, as observed in our previous work, are capable of displaying immiscible behaviour at high lengths and rigidities, despite a small effective size due to the lack of side chains. An effort was made in this work to include such domain knowledge into the model creation by force sampling all possible zero side chain PL molecules across backbone lengths and flexibilities of interest (see Supporting Information). Despite this, the model predicts these molecules as miscible. This is most likely because, even with such measures, this region of the feature space is heavily underrepresented and so the model is unable to learn the correct behaviour. \par

\begin{figure}[H] 
    \centering
    \includegraphics[width=0.95\textwidth]{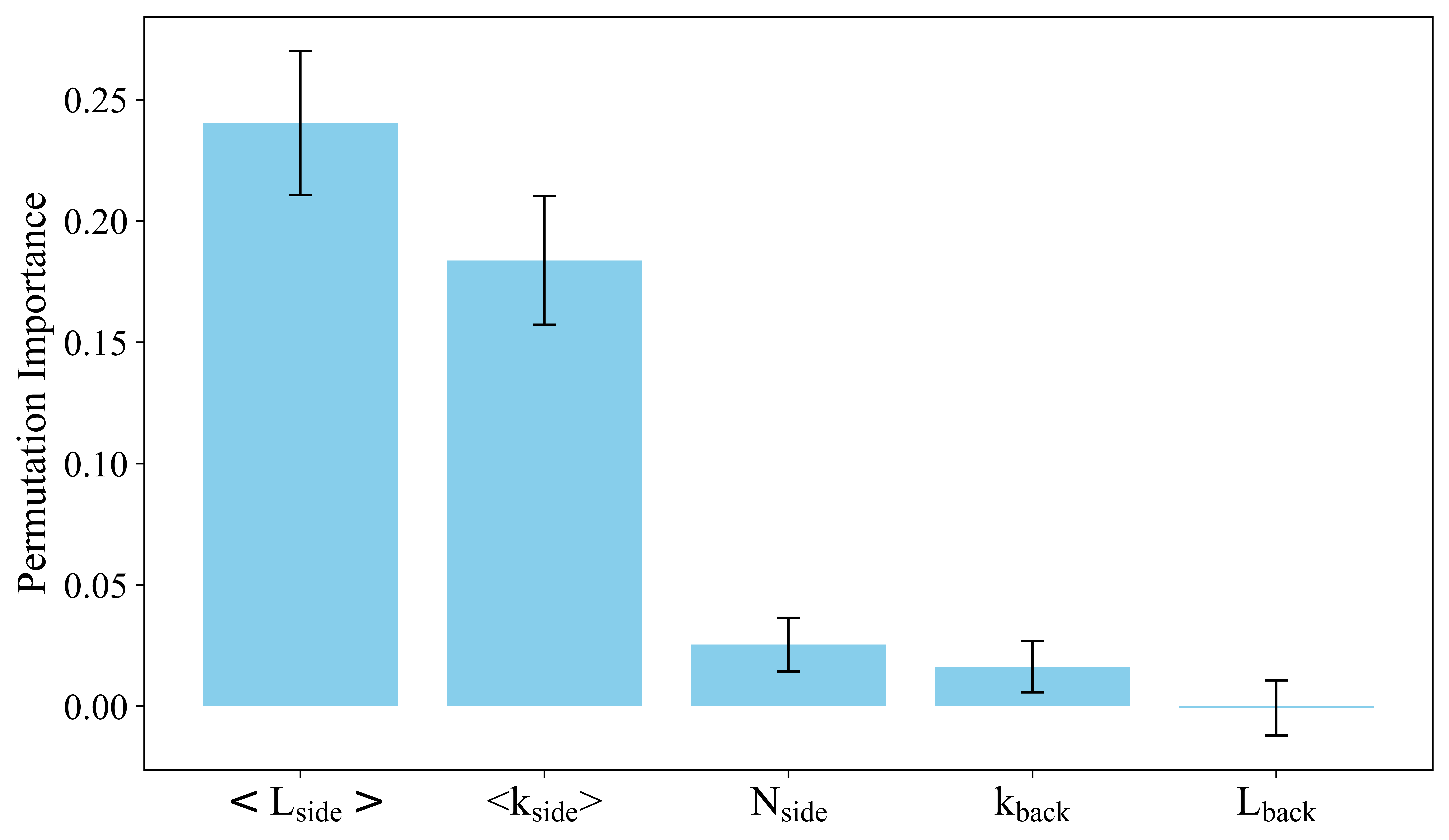} 
    \captionsetup{font={stretch=1.5}}
    \caption{Permutation importance, $I_j$, for PL descriptors average side chain length, $<L_{side}>$, average side chain flexibility, $<k_{side}>$, side chain number, $N_{side}$, backbone flexibility, $k_{back}$ and backbone length, $L_{back}$. Error bars are the standard deviation of $I_j$ over 30 shuffles of each feature.}
    \label{fig:perm} 
\end{figure}

\noindent The permutation importance, $I_j$, of the model features can provide us with a final insight into the PL features which are the most informative when determining their miscibility. To calculate this, the feature of interest is randomly shuffled within the hold-out test set, $D$ to form a corrupted dataset, $\tilde{D}_{j,k}$. A baseline performance metric, $s$, (F1 score) is extracted from the original data and the permutation importance for the feature is computed with

\begin{equation}
    I_j = s - \frac{1}{K}\sum_{k=1}^K \tilde{s}_{k,j}
\end{equation}

\noindent where $K$ is the number of shuffle repetitions and $\tilde{s}_{k,j}$ is the F1 score of the corrupted dataset. $I_j$ then represents the feature importance measured by the decrease in model performance upon its permutation. Figure~\ref{fig:perm} displays $I_j$ for each PL descriptor used in the RF model.

\noindent From the results, it is clear that two features dominate the decisions made within the RF model, $<L_{side}>$ and $<k_{side}>$. The other features, $N_{side}$, $k_{back}$ and $L_{back}$, have $I_j < 0.05$ and so contribute minimally. We may justify this result by considering the format of our data. In the case of $<L_{side}>$, high values of $<L_{side}>$ must correspond to a value of $L_{back}$ at least one unit higher. Conversely, a high $L_{back}$ does not require a high $<L_{side}>$, thus $L_{back}$ becomes a redundant feature. Therefore, the high importance of $L_{side}$ is further confirmation that the bulkiness of the PL molecules is a primary feature for predicting its miscibility. \par

\noindent The second most important feature, $<k_{side}>$, can be rationalised by considering that, in the majority of our data, $<k_{side}> = k_{back}$, and so the value of $<k_{side}>$ is indicative to the flexibility of the whole molecule, which is an informative feature for deciding PL miscibility. The rest of the data consists of molecules which have either opposite backbone and side chain flexibilities or a rigid backbone with alternating rigid/flexible side chains. In such cases, as we previously discussed, $<k_{side}>$ plays an important role in the bulkiness of the PL molecule and, consequently, PL miscibility. \par

\noindent These findings re-affirm our findings that the PL molecule bulkiness and side chain flexibility plays a primary role in their miscibility behaviour. They can also be used to inform future expansions of this work, for instance the investigation of a variety of polyolefin melts mixed with non-polar plasticisers, for which we can now consider removing features which likely have low permutation importance and thereby reducing the surrogate model dimensionality.

\section{Summary and Conclusions}

\noindent In this work, we have demonstrated that, using a coarse-grained bead-and-spring model, we are able to predict the phase behaviour of generic PL molecules of varying degrees of rigidities and bulkiness using a RF surrogate model. Due to the high cost associated with obtaining the label for phase separation (miscible = 0, immiscible = 1) with a PI/PL MD simulation, we build the model using a pool-based active learning procedure with a diverse unlabelled pool spanning 2055 PLs with varying backbone lengths, average side chain lengths, side chain numbers, backbone flexibilities and average side chain flexibilities. We envision that such PLs represent non-polar molecules of varying degrees of inter- and intra- saturation, branching and molecular weights. \par

\noindent Using an uncertainty/random hybrid query strategy which balances design space exploitation and exploration, we iteratively label 108 PL molecules over 8 AL rounds and achieve a surrogate model F1 score of 0.89, which we use to make predictions within our design space. We find that rigid PLs with bulky side chains are more likely to be immiscible compared to their flexible, smaller counterparts. Additionally, those PLs with rigid backbones but flexible side chains are likely to be miscible, whereas the opposite case, of molecules with a flexible backbone but long, rigid side chains, are likely to be immiscible. Thus, we can infer that side chain rigidity plays a larger role in PI/PL phase behaviour compared to backbone rigidity. \par

\noindent This work has allowed us to isolate the impact of tailored PL features to explore their impact on PI/PL miscibility, which could feasibly be used in the design of new PLs for industrial purposes. To support this conclusion, we also validated PL phase behaviour for 6 atomistic hydrocarbons in the absence of side chains, both with PI/PL simulations and Hildebrand solubility parameters. We found that the PI/PL phase behaviour in these systems agreed with that of the coarse grained model and, in the future, further validation for those PL molecules with different degrees of branching will be performed.  \par 

\noindent Overall, this work makes significant progress towards a multi-scale workflow in which generic PL design rules can be used to inform the design of PLs with finer resolution. Thus, an initial screening which eliminates unfavourable areas of the design space can be performed without large numbers of costly, atomistic simulations.  

\clearpage

\begin{center}
\Large 
\textbf{Supporting Information for} \\
\huge
Active Learning for Predicting Polymer/Plasticizer Phase Behaviour\\ 

\hspace{10pt}

\small
Lois Smith$^1$, Jessica Steele$^1$, Hossein Ali Karimi-Varzaneh$^2$, Paola Carbone$^1$ \\

\hspace{10pt}

\small  
$^1$ Department of Chemistry, School of Natural Sciences, The University of Manchester, M13 9PL, Manchester, United Kingdom \\
$^2$ Continental Reifen, Deutschland GmbH, D-30419 Hanover, Germany \\
\end{center} 

\clearpage

\noindent \textit{\large{Unlabelled Pool Sampling}} \par

\noindent To create the PL design space, we generated PL molecules with all allowed combinations of flexibility, side chain placements and side chain lengths, along backbone lengths of 4 to 14, the requirements for which are provided in \textit{Coarse-Grained PLs} of the main text. PLs with mirror symmetry were removed as they are equivalent. The total combinatorial space spanned over 456 million points. We then randomly sampled approximately 2 million points from the space. \par

\noindent To ensure we captured the original data sparsity, whilst also maintaining underrepresented areas of the space, such as those with small values of $L_{back}$, we ensured that the relative proportions of molecules with large ($L_{back}=12-14$ beads: $\approx$ 99.4$\%$), mid ($L_{back}=8-11$ beads: $\approx$ 0.57$\%$), and small backbone lengths ($L_{back}=4-7$ beads: $\approx$ 0.001$\%$) remained approximately consistent between the down-sampled and original PL design spaces. However, since this amounts to around 20 PLs with small backbone lengths, we made an adjustment to increase the proportion of such molecules to $0.02\%$ of the sampled space. This increased their number to 394. \par

\noindent To create a PL design space representing diversity in the core features of interest: $L_{back}$, $<L_{side}>$, $N_{side}$, $k_{back}$ and $<k_{side}>$, we employed a diversity sampling procedure to sample 2055 points, which we believe is a sufficient amount to capture a range of PL sizes and flexibilities. \par

\noindent The procedure begins with the selection of a random point in the search space. Then, the pairwise Euclidean distances between this point and all other points in the space was computed, and the point with the maximum distance was added to the sample. In the next cycle, pairwise distances are also calculated for the new point and the maximin point is added to the sample. \par

\noindent From our previous work,~\cite{smith2024framework} which signifies that larger and more rigid PL molecules tend to be immiscible in the PI melt, we suspect a highly imbalanced dataset. Larger PL molecules (higher $L_{back}$ and $<L_{side}>$) also naturally span a larger combinatorial space in terms of their allowed side chain positions and lengths. To preserve this in the dataset, and to safeguard against the majority class (miscible=0) dominating the PL design space, we applied a scaling factor of 3.5 to the pairwise distance calculations to PLs with a $L_{back}\leq10$, $<L_{side}>\geq 7$ and whole molecule flexibility corresponding to a value of $k_h$ between $70$ and $100~\text{kJ/mol}$. The resulting number of PLs satisfying these conditions is 821, approximately $41\%$ of the design space. \par

\noindent Finally, since PL molecules with no side chains are an area of particular interest, we chose to force sample every possible configuration within the allowed values of $L_{back}$ and $k_{back}$ in order to maximise the size of this design sub-space. This was done due to the findings of our previous work~\cite{RN116} which found rod-like PLs of a high molecular weight and rigidity are immiscible within the PI melt, despite having a smaller effective size than molecules of the same $L_{back}$ and $k_{back}$, but with the addition of small side chains. \par

\begin{figure}[htp]
\centering

\begin{subfigure}{0.49\columnwidth}
\centering
\includegraphics[width=\textwidth]{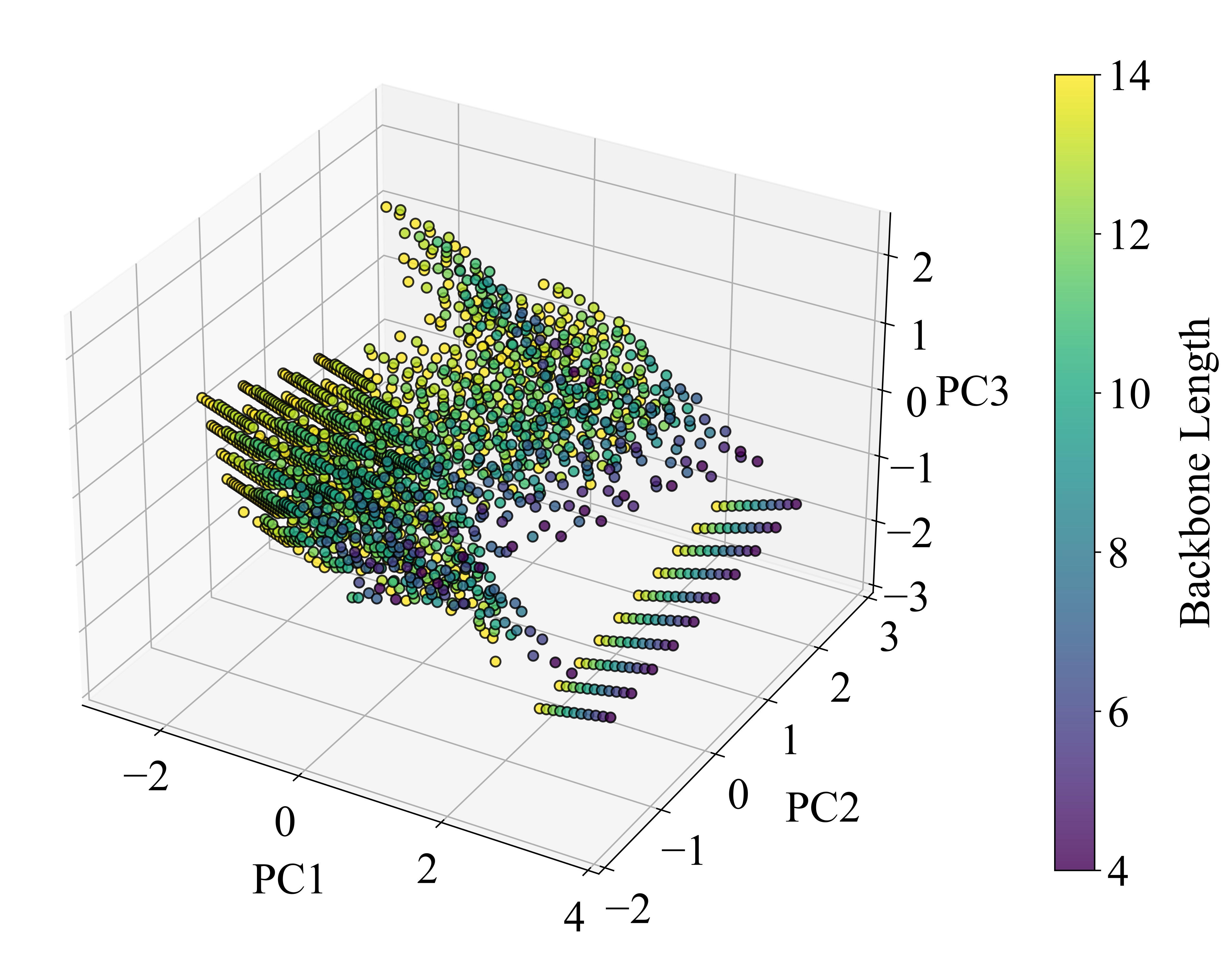}
\caption{}
\label{fig:time1}
\end{subfigure}\hfill
\begin{subfigure}{0.49\columnwidth}
\centering
\includegraphics[width=\textwidth]{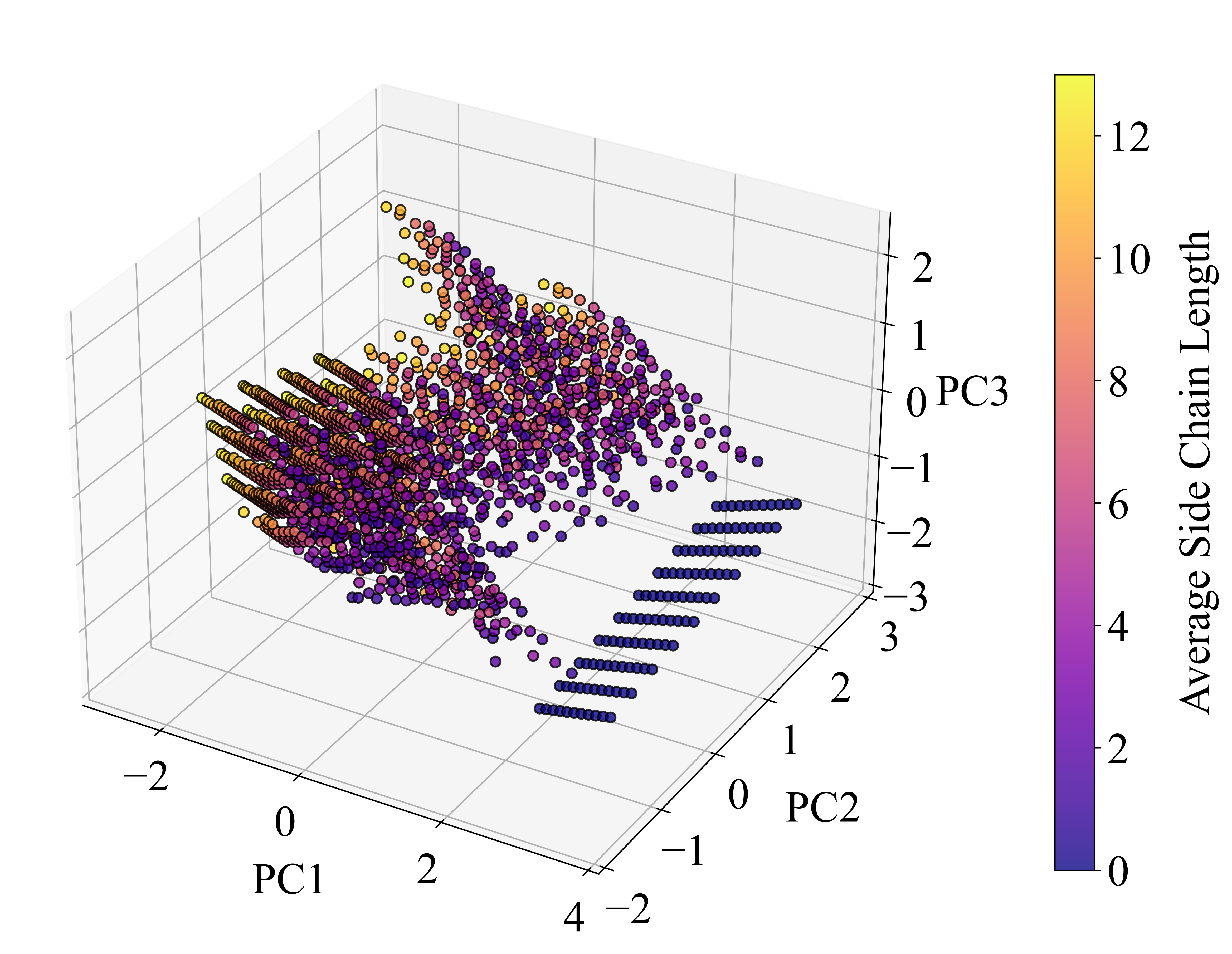}
\caption{}
\label{fig:time2}
\end{subfigure}

\medskip

\begin{subfigure}{0.49\columnwidth}
\centering
\includegraphics[width=\textwidth]{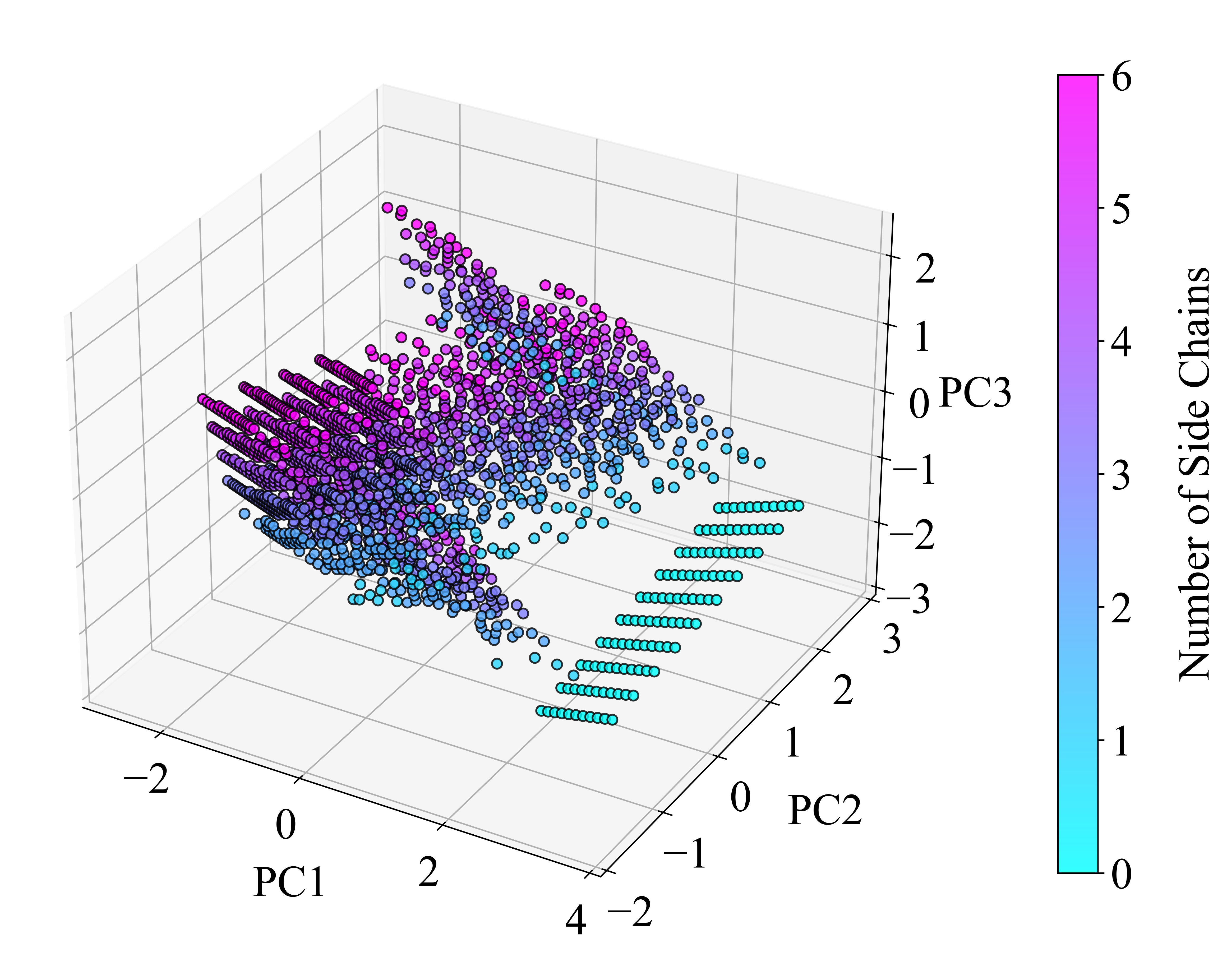}
\caption{}
\label{fig:time3}
\end{subfigure}\hfill
\begin{subfigure}{0.49\columnwidth}
\centering
\includegraphics[width=\textwidth]{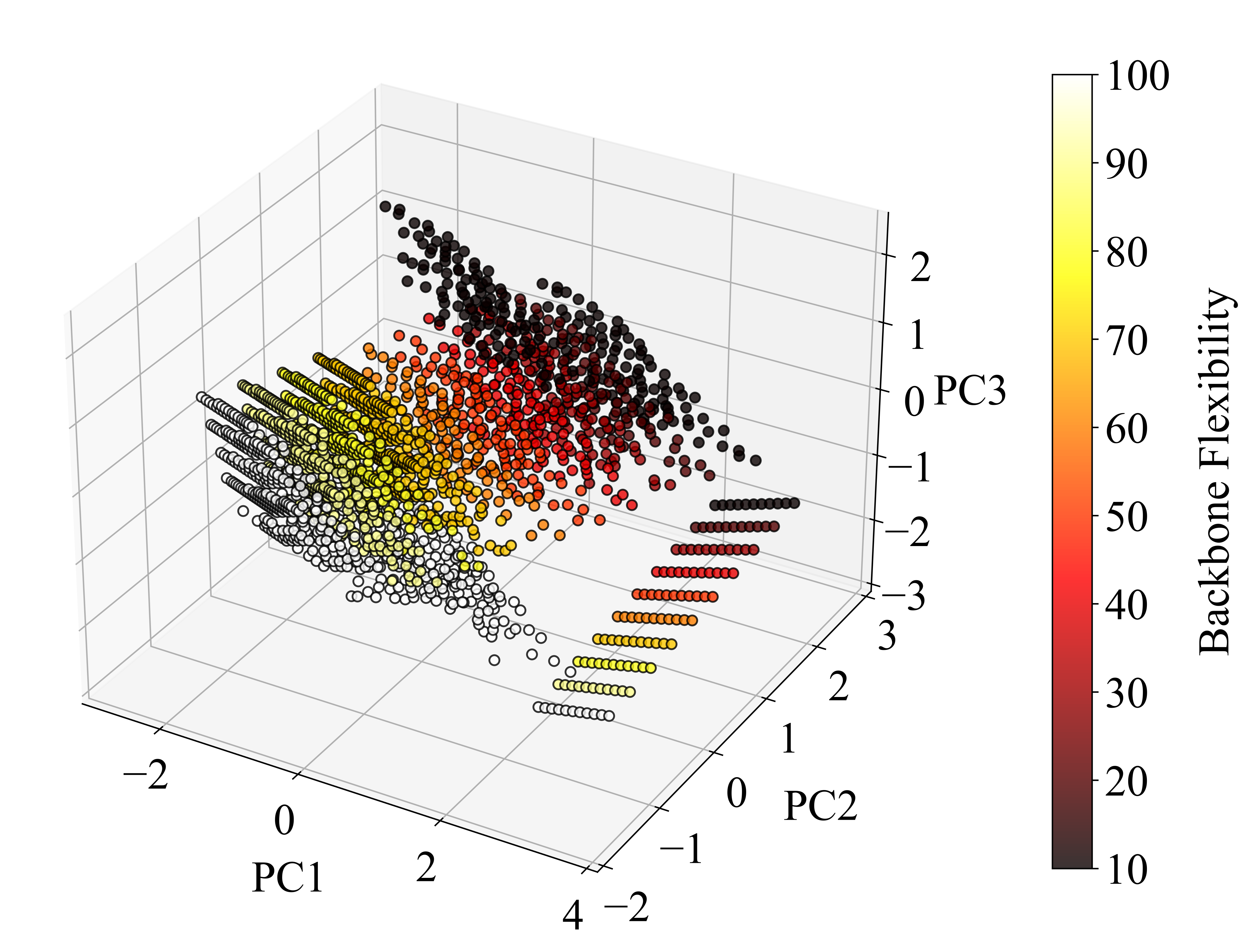}
\caption{}
\label{fig:time4}
\end{subfigure}

\medskip

\begin{subfigure}{0.49\columnwidth}
\centering
\includegraphics[width=\textwidth]{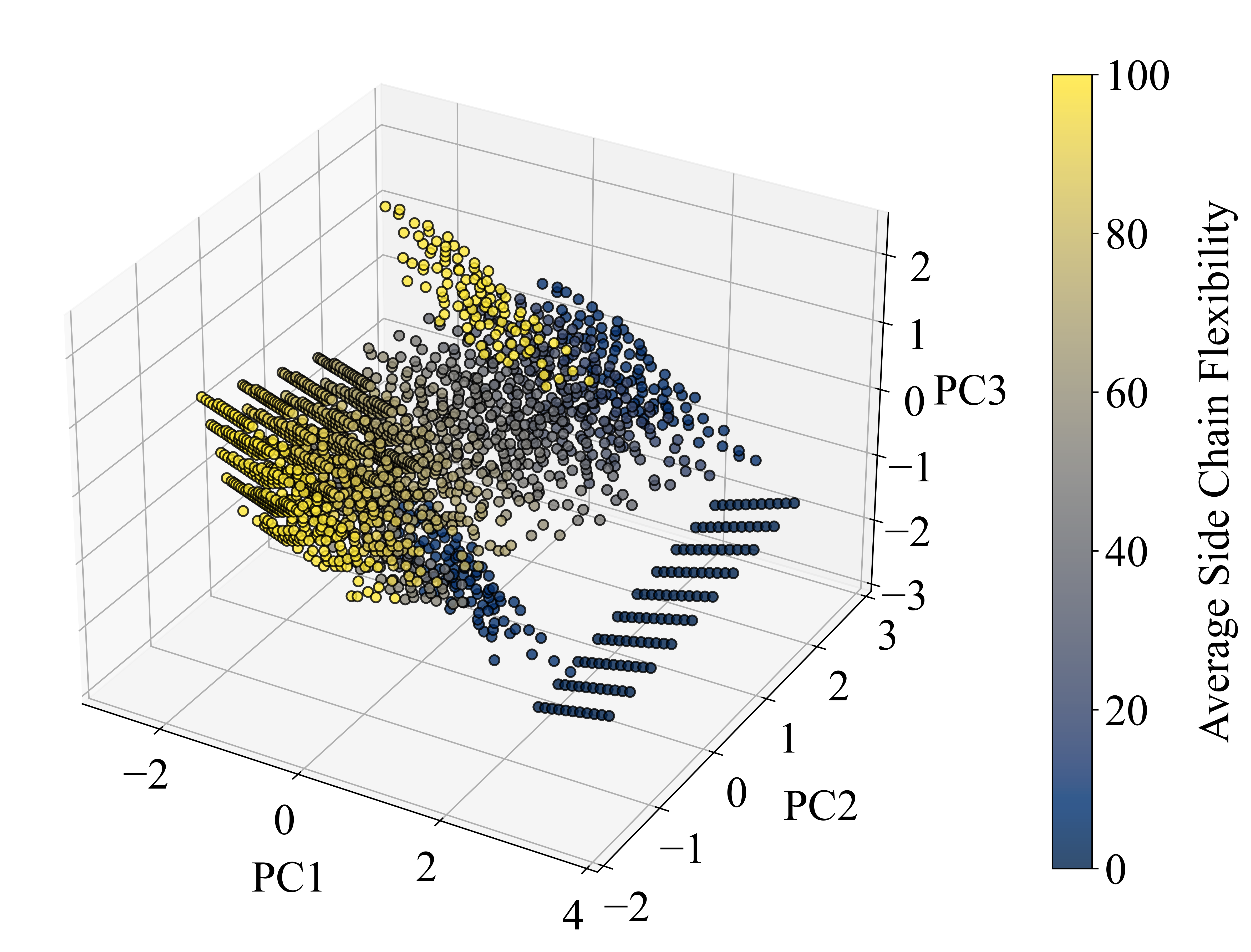}
\caption{}
\label{fig:time5}
\end{subfigure}
\captionsetup{font={stretch=1.5}}
\caption{Display of the initial unlabelled data space projected into 3 principal components. Plots are coloured by (a) backbone length, (b) average side chain length, (c) side chain number, (d) backbone flexibility and (e) average side chain flexibility.}
\label{fig:pca-sample}

\end{figure}

\noindent The final sampled space is formed of 2055 PL molecules and is visualised in 3-dimensions using PCA in Figure~\ref{fig:pca-sample}. Sub-plots are coloured according to the PL features from which the space was sampled: $L_{back}$, $N_{side}$, $<L_{side}>$, $k_{back}$ and $<k_{side}>$. Most noticeably, we can observe an ordered region of points, separated from the bulk of the data, which represents PLs without side chains. They have values of $N_{side}$, $<L_{side}>$ and $<k_{side}>=0$, unlike all other data points, which justifies their unusual projections onto the principal components. \par

\noindent The remainder of the data covers all PL feature ranges of interest, with an expected bias towards larger PLs, which is natural in the data. Particularly though, Figures~\ref{fig:pca-sample}(d) and (e) also highlight well areas of the design space which have differing values of $k_{back}$ and $<k_{side}>$, displayed on the `points' of data on the lower and upper parts of the plots. Thus, our sampling procedure accurately captures all important regions from the original PL design space.

\clearpage


\bibliographystyle{unsrt}
\bibliography{bibliography.bib}

\begin{thebibliography}{10}

\bibitem{ROBESON1999549}
L.M Robeson.
\newblock Polymer membranes for gas separation.
\newblock {\em Current Opinion in Solid State and Materials Science}, 4(6):549--552, 1999.

\bibitem{Lasseuguette2022-zy}
Elsa Lasseuguette and Bibiana Comesa{\~n}a-G{\'a}ndara.
\newblock Polymer membranes for gas separation.
\newblock {\em Membranes (Basel)}, 12(2), February 2022.

\bibitem{SIDHIKKUKANDATHVALAPPIL2021103}
Riya {Sidhikku Kandath Valappil}, Nayef Ghasem, and Mohamed Al-Marzouqi.
\newblock Current and future trends in polymer membrane-based gas separation technology: A comprehensive review.
\newblock {\em Journal of Industrial and Engineering Chemistry}, 98:103--129, 2021.

\bibitem{Oyaizu2024}
Kenichi Oyaizu.
\newblock Reversible and high-density energy storage with polymers populated with bistable redox sites.
\newblock {\em Polymer Journal}, 56(3):127--144, Mar 2024.

\bibitem{LI2023101714}
Chuanfa Li, Kun Zhang, Xiangran Cheng, Jiaxin Li, Yi~Jiang, Pengzhou Li, Bingjie Wang, and Huisheng Peng.
\newblock Polymers for flexible energy storage devices.
\newblock {\em Progress in Polymer Science}, 143:101714, 2023.

\bibitem{C4RA15947K}
Muhammad~E. Abdelhamid, Anthony~P. O{'}Mullane, and Graeme~A. Snook.
\newblock Storing energy in plastics: a review on conducting polymers \& their role in electrochemical energy storage.
\newblock {\em RSC Adv.}, 5:11611--11626, 2015.

\bibitem{TAJEDDIN2020525}
Behjat Tajeddin and Mina Arabkhedri.
\newblock Chapter 16 - polymers and food packaging.
\newblock In Mariam Al~Ali AlMaadeed, Deepalekshmi Ponnamma, and Marcelo~A. Carignano, editors, {\em Polymer Science and Innovative Applications}, pages 525--543. Elsevier, 2020.

\bibitem{Patil2017}
Akshat Patil, Arun Patel, and Rajesh Purohit.
\newblock An overview of polymeric materials for automotive applications.
\newblock {\em Materials Today: Proceedings}, 4(2, Part A):3807--3815, Jan 2017.

\bibitem{YUE2022107584}
Tongkui Yue, Zhiyu Zhang, Sai Li, Hengheng Zhao, Pengwei Duan, Ganggang Zhang, Liqun Zhang, and Jun Liu.
\newblock Designing and fabricating nanopolymer composites beyond traditional polymer nanocomposites toward fuel saving of automobile tires.
\newblock {\em Nano Energy}, 101:107584, 2022.

\bibitem{doi:10.1021/ma9814548}
Benny~D. Freeman.
\newblock Basis of permeability/selectivity tradeoff relations in polymeric gas separation membranes.
\newblock {\em Macromolecules}, 32(2):375--380, 1999.

\bibitem{Deshpande2014}
Pravin~P. Deshpande, Niteen~G. Jadhav, Victoria~J. Gelling, and Dimitra Sazou.
\newblock Conducting polymers for corrosion protection: a review.
\newblock {\em Journal of Coatings Technology and Research}, 11(4):473--494, Jul 2014.

\bibitem{li2023ionic}
Zhuo Li, Jialong Fu, Xiaoyan Zhou, Siwei Gui, Lu~Wei, Hui Yang, Hong Li, and Xin Guo.
\newblock Ionic conduction in polymer-based solid electrolytes.
\newblock {\em Advanced Science}, 10(10):2201718, 2023.

\bibitem{parin2024durability}
Fatma~Nur PARIN and Fatma DEMIRCI.
\newblock Durability of polymer composite materials for high-temperature applications.
\newblock In {\em Aging and Durability of FRP Composites and Nanocomposites}, pages 135--170. Elsevier, 2024.

\bibitem{Li}
Ying Li and Tianle Yue.
\newblock {\em Machine Learning for Polymer Informatics}.
\newblock American Chemical Society, Washington, DC, USA, 2024.

\bibitem{CHEN2021100595}
Lihua Chen, Ghanshyam Pilania, Rohit Batra, Tran~Doan Huan, Chiho Kim, Christopher Kuenneth, and Rampi Ramprasad.
\newblock Polymer informatics: Current status and critical next steps.
\newblock {\em Materials Science and Engineering: R: Reports}, 144:100595, 2021.

\bibitem{audus}
Debra~J. Audus and Juan~J. de~Pablo.
\newblock Polymer informatics: Opportunities and challenges.
\newblock {\em ACS Macro Letters}, 6(10):1078--1082, 2017.

\bibitem{Struble2024}
Daniel~C. Struble, Bradley~G. Lamb, and Boran Ma.
\newblock A prospective on machine learning challenges, progress, and potential in polymer science.
\newblock {\em MRS Communications}, 14(5):752--770, Oct 2024.

\bibitem{tao2021machine}
Lei Tao, Guang Chen, and Ying Li.
\newblock Machine learning discovery of high-temperature polymers.
\newblock {\em Patterns}, 2(4), 2021.

\bibitem{park2022prediction}
Jaehong Park, Youngseon Shim, Franklin Lee, Aravind Rammohan, Sushmit Goyal, Munbo Shim, Changwook Jeong, and Dae~Sin Kim.
\newblock Prediction and interpretation of polymer properties using the graph convolutional network.
\newblock {\em ACS Polymers Au}, 2(4):213--222, 2022.

\bibitem{genome}
Chiho Kim, Anand Chandrasekaran, Tran~Doan Huan, Deya Das, and Rampi Ramprasad.
\newblock Polymer genome: A data-powered polymer informatics platform for property predictions.
\newblock {\em The Journal of Physical Chemistry C}, 122(31):17575--17585, 2018.

\bibitem{casanola2024machine}
Gerardo~M Casanola-Martin, Anas Karuth, Hai Pham-The, Humbert Gonz{\'a}lez-D{\'\i}az, Dean~C Webster, and Bakhtiyor Rasulev.
\newblock Machine learning analysis of a large set of homopolymers to predict glass transition temperatures.
\newblock {\em Communications Chemistry}, 7(1):226, 2024.

\bibitem{wu2019machine}
Stephen Wu, Yukiko Kondo, Masa-aki Kakimoto, Bin Yang, Hironao Yamada, Isao Kuwajima, Guillaume Lambard, Kenta Hongo, Yibin Xu, Junichiro Shiomi, et~al.
\newblock Machine-learning-assisted discovery of polymers with high thermal conductivity using a molecular design algorithm.
\newblock {\em Npj Computational Materials}, 5(1):66, 2019.

\bibitem{huang2023exploring}
Xiang Huang, Shengluo Ma, CY~Zhao, Hong Wang, and Shenghong Ju.
\newblock Exploring high thermal conductivity polymers via interpretable machine learning with physical descriptors.
\newblock {\em npj Computational Materials}, 9(1):191, 2023.

\bibitem{chen2020frequency}
Lihua Chen, Chiho Kim, Rohit Batra, Jordan~P Lightstone, Chao Wu, Zongze Li, Ajinkya~A Deshmukh, Yifei Wang, Huan~D Tran, Priya Vashishta, et~al.
\newblock Frequency-dependent dielectric constant prediction of polymers using machine learning.
\newblock {\em npj Computational Materials}, 6(1):61, 2020.

\bibitem{liang2021machine}
Jiechun Liang, Shangqian Xu, Linfeng Hu, Yu~Zhao, and Xi~Zhu.
\newblock Machine-learning-assisted low dielectric constant polymer discovery.
\newblock {\em Materials Chemistry Frontiers}, 5(10):3823--3829, 2021.

\bibitem{xu2021machine}
Pengcheng Xu, Tian Lu, Lifei Ju, Lumin Tian, Minjie Li, and Wencong Lu.
\newblock Machine learning aided design of polymer with targeted band gap based on dft computation.
\newblock {\em The Journal of Physical Chemistry B}, 125(2):601--611, 2021.

\bibitem{alzahrani2024machine}
Fatimah Mohammed~A Alzahrani, Muhammad Sagir, Muhammad Saqib, Shahida Bashir, Tayyaba Sarwar, Shabbir Hussain, Shahzad Murtaza, Afifa Mushtaq, Rafia Razzaq, ZA~Alrowaili, et~al.
\newblock Machine learning assisted prediction of band gaps and designing of new polymers for photodetectors: A complete pipeline.
\newblock {\em Computational Materials Science}, 239:112961, 2024.

\bibitem{lightstone2020refractive}
Jordan~P Lightstone, Lihua Chen, Chiho Kim, Rohit Batra, and Rampi Ramprasad.
\newblock Refractive index prediction models for polymers using machine learning.
\newblock {\em Journal of Applied Physics}, 127(21), 2020.

\bibitem{sivaraman2020machine}
Ganesh Sivaraman, Nicholas~E Jackson, Benjamin Sanchez-Lengeling, {\'A}lvaro V{\'a}zquez-Mayagoitia, Al{\'a}n Aspuru-Guzik, Venkatram Vishwanath, and Juan~J De~Pablo.
\newblock A machine learning workflow for molecular analysis: application to melting points.
\newblock {\em Machine Learning: Science and Technology}, 1(2):025015, 2020.

\bibitem{Jiang2024}
Shengli Jiang, Adji~Bousso Dieng, and Michael~A. Webb.
\newblock Property-guided generation of complex polymer topologies using variational autoencoders.
\newblock {\em npj Computational Materials}, 10(1):139, Jun 2024.

\bibitem{ethier2023integrating}
Jeffrey~G Ethier, Debra~J Audus, Devin~C Ryan, and Richard~A Vaia.
\newblock Integrating theory with machine learning for predicting polymer solution phase behavior.
\newblock {\em Giant}, 15:100171, 2023.

\bibitem{gentekos2019controlling}
Dillon~T Gentekos, Renee~J Sifri, and Brett~P Fors.
\newblock Controlling polymer properties through the shape of the molecular-weight distribution.
\newblock {\em Nature Reviews Materials}, 4(12):761--774, 2019.

\bibitem{yang2023identification}
Peilin Yang, Wei Gao, and Eric Wasserman.
\newblock Identification and quantification of polymeric impurity in block copolymer by one-dimensional and two-dimensional liquid chromatography coupled to high-resolution mass spectrometry and evaporative light scattering detector.
\newblock {\em Journal of Chromatography A}, 1694:463909, 2023.

\bibitem{chandran2019processing}
Sivasurender Chandran, J{\"o}rg Baschnagel, Daniele Cangialosi, Koji Fukao, Emmanouil Glynos, Liesbeth~MC Janssen, Marcus M{\"u}ller, Murugappan Muthukumar, Ullrich Steiner, Jun Xu, et~al.
\newblock Processing pathways decide polymer properties at the molecular level.
\newblock {\em Macromolecules}, 52(19):7146--7156, 2019.

\bibitem{AMBROGI201787}
V.~Ambrogi, C.~Carfagna, P.~Cerruti, and V.~Marturano.
\newblock 4 - additives in polymers.
\newblock In Carlos~F. Jasso-Gastinel and José~M. Kenny, editors, {\em Modification of Polymer Properties}, pages 87--108. William Andrew Publishing, 2017.

\bibitem{RN78}
Kevin~Maik Jablonka, Giriprasad~Melpatti Jothiappan, Shefang Wang, Berend Smit, and Brian Yoo.
\newblock Bias free multiobjective active learning for materials design and discovery.
\newblock {\em Nature Communications}, 12(1):2312, 2021.

\bibitem{polyinfo}
Shingo Otsuka, Isao Kuwajima, Junko Hosoya, Yibin Xu, and Masayoshi Yamazaki.
\newblock Polyinfo: Polymer database for polymeric materials design.
\newblock In {\em 2011 International Conference on Emerging Intelligent Data and Web Technologies}, pages 22--29, 2011.

\bibitem{SHETTY2021101922}
Pranav Shetty and Rampi Ramprasad.
\newblock Automated knowledge extraction from polymer literature using natural language processing.
\newblock {\em iScience}, 24(1):101922, 2021.

\bibitem{luo2024machine}
Gaoyang Luo, Feicheng Huan, Yuwei Sun, Feng Shi, Shengwei Deng, and Jian-guo Wang.
\newblock Machine learning-based high-throughput screening for high-stability polyimides.
\newblock {\em Industrial \& Engineering Chemistry Research}, 63(48):21110--21122, 2024.

\bibitem{marti}
Didac Martí, R{\'e}mi P{\'e}tuya, Emanuele Bosoni, Anne-Claude Dublanchet, Stephan Mohr, and Fabien L{\'e}onforte.
\newblock Predicting the glass transition temperature of biopolymers via high-throughput molecular dynamics simulations and machine learning.
\newblock {\em ACS Applied Polymer Materials}, 6(8):4449--4461, 2024.

\bibitem{MA2022100850}
Ruimin Ma, Hanfeng Zhang, Jiaxin Xu, Luning Sun, Yoshihiro Hayashi, Ryo Yoshida, Junichiro Shiomi, Jian xun Wang, and Tengfei Luo.
\newblock Machine learning-assisted exploration of thermally conductive polymers based on high-throughput molecular dynamics simulations.
\newblock {\em Materials Today Physics}, 28:100850, 2022.

\bibitem{nanjo2025spacier}
Shun Nanjo, Arifin, Hayato Maeda, Yoshihiro Hayashi, Kan Hatakeyama-Sato, Ryoji Himeno, Teruaki Hayakawa, and Ryo Yoshida.
\newblock Spacier: on-demand polymer design with fully automated all-atom classical molecular dynamics integrated into machine learning pipelines.
\newblock {\em npj Computational Materials}, 11(1):16, 2025.

\bibitem{martini2018review}
Ashlie Martini, Uma~Shantini Ramasamy, and Michelle Len.
\newblock Review of viscosity modifier lubricant additives.
\newblock {\em Tribology Letters}, 66:1--14, 2018.

\bibitem{van2021role}
Bas~GP Van~Ravensteijn, Raghida~Bou Zerdan, Craig~J Hawker, and Matthew~E Helgeson.
\newblock Role of architecture on thermorheological properties of poly (alkyl methacrylate)-based polymers.
\newblock {\em Macromolecules}, 54(12):5473--5483, 2021.

\bibitem{adeyemi2022equilibrium}
Oluseye Adeyemi, Shiping Zhu, and Li~Xi.
\newblock Equilibrium and non-equilibrium molecular dynamics approaches for the linear viscoelasticity of polymer melts.
\newblock {\em Physics of Fluids}, 34(5), 2022.

\bibitem{critical}
Shruti Venkatram, Chiho Kim, Anand Chandrasekaran, and Rampi Ramprasad.
\newblock Critical assessment of the hildebrand and hansen solubility parameters for polymers.
\newblock {\em Journal of Chemical Information and Modeling}, 59(10):4188--4194, 2019.

\bibitem{roadmap}
Thomas E.~III Gartner and Arthi Jayaraman.
\newblock Modeling and simulations of polymers: A roadmap.
\newblock {\em Macromolecules}, 52(3):755--786, 2019.

\bibitem{RN52}
O.~A. Al-Hartomy, F.~Al-Solamy, A.~Al-Ghamdi, N.~Dishovsky, M.~Ivanov, M.~Mihaylov, and F.~El-Tantawy.
\newblock Influence of carbon black structure and specific surface area on the mechanical and dielectric properties of filled rubber composites.
\newblock {\em International Journal of Polymer Science}, 2011:33--39, 2011.

\bibitem{RN53}
C.~G. Robertson and N.~J. Hardman.
\newblock Nature of carbon black reinforcement of rubber: Perspective on the original polymer nanocomposite.
\newblock {\em Polymers (Basel)}, 13(4):577--594, 2021.

\bibitem{Edwards1990}
D.~C. Edwards.
\newblock Polymer-filler interactions in rubber reinforcement.
\newblock {\em Journal of Materials Science}, 25(10):4175--4185, Oct 1990.

\bibitem{RN51}
S.~S. Sarkawi, W.~K. Dierkes, and J.~W.~M. Noordermeer.
\newblock Elucidation of filler-to-filler and filler-to-rubber interactions in silica-reinforced natural rubber by tem network visualization.
\newblock {\em European Polymer Journal}, 54:118--127, 2014.

\bibitem{RN56}
S.~Salina Sarkawi, Wilma~K. Dierkes, and Jacques W.~M. Noordermeer.
\newblock Morphology of silica-reinforced natural rubber: The effect of silane coupling agent.
\newblock {\em Rubber Chemistry and Technology}, 88(3):359--372, 2015.

\bibitem{RN47}
S.~S. Choi, C.~Nah, S.~G. Lee, and C.~W. Joo.
\newblock Effect of filler-filler interaction on rheological behaviour of natural rubber compounds filled with both carbon black and silica.
\newblock {\em Polymer International}, 52(1):23--28, 2003.

\bibitem{wolff1992filler}
Siegfried Wolff and Meng-Jiao Wang.
\newblock Filler—elastomer interactions. part iv. the effect of the surface energies of fillers on elastomer reinforcement.
\newblock {\em Rubber chemistry and technology}, 65(2):329--342, 1992.

\bibitem{RN111}
Jaeho Oh, Yong~Hwan Yoo, Il-Sou Yoo, Yang-Il Huh, Tapan~Kumar Chaki, and Changwoon Nah.
\newblock Effect of plasticizer and curing system on freezing resistance of rubbers.
\newblock {\em Journal of Applied Polymer Science}, 131(2):39795--39803, 2014.

\bibitem{RN115}
C.~Bergmann and J.~Trimbach.
\newblock Influence of plasticizers on the properties of natural rubber based compounds.
\newblock {\em KGK-Kautschuk Gummi Kunststoffe}, 67(7-8):40--49, 2014.

\bibitem{RN57}
C.S. Brazel and S.L. Rosen.
\newblock {\em Fundamental Principles of Polymeric Materials, 3rd Edition}.
\newblock John Wiley and Sons, 2012.

\bibitem{RN99}
Hüsamettin~D. Özeren, Marcel Balçık, M.~Göktuǧ Ahunbay, and J.~Richard Elliott.
\newblock In silico screening of green plasticizers for poly(vinyl chloride).
\newblock {\em Macromolecules}, 52(6):2421--2430, 2019.

\bibitem{RN101}
Basant~M. Elsiwi, Omar Garcia-Valdez, Hanno~C. Erythropel, Richard~L. Leask, Jim~A. Nicell, and Milan Maric.
\newblock Fully renewable, effective, and highly biodegradable plasticizer: Di-n-heptyl succinate.
\newblock {\em ACS Sustainable Chemistry \& Engineering}, 8(33):12409--12418, 2020.

\bibitem{RN112}
Nasruddin and T.~Susanto.
\newblock The effect of natural based oil as plasticizer towards physics-mechanical properties of nr-sbr blending for solid tyres, 2018.

\bibitem{RN113}
J.~Thomas and R.~Patil.
\newblock The road to sustainable tire materials: Current state-of-the-art and future prospectives.
\newblock {\em Environmental Science \& Technology}, 57(6):2209--2216, 2023.

\bibitem{RN114}
W.~K. Gao, W.~J. Zhang, J.~H. Liu, and J.~Hua.
\newblock Multifunctional and anti-migration plasticizers based on bio-based phenol-functionalized liquid 1,2-polybutadiene.
\newblock {\em Polymer Testing}, 118, 2023.

\bibitem{RN72}
C.~Y. Bao, D.~R. Long, and C.~Vergelati.
\newblock Miscibility and dynamical properties of cellulose acetate/plasticizer systems.
\newblock {\em Carbohydrate Polymers}, 116:95--102, 2015.

\bibitem{RN71}
N.~Lindemann, S.~Finger, H.~A. Karimi-Varzaneh, and J.~Lacayo-Pineda.
\newblock Rigidity of plasticizers and their miscibility in silica-filled polybutadiene rubber by broadband dielectric spectroscopy.
\newblock {\em Journal of Applied Polymer Science}, 139(21), 2022.

\bibitem{RN69}
A.~Jarray, V.~Gerbaud, and M.~Hemati.
\newblock Polymer-plasticizer compatibility during coating formulation: A multi-scale investigation.
\newblock {\em Progress in Organic Coatings}, 101:195--206, 2016.

\bibitem{RN121}
Antonio Greco, Francesca Ferrari, and Alfonso Maffezzoli.
\newblock Mechanical properties of poly(lactid acid) plasticized by cardanol derivatives.
\newblock {\em Polymer Degradation and Stability}, 159:199--204, 2019.

\bibitem{smith2024framework}
Lois Smith, Hossein~Ali Karimi-Varzaneh, Sebastian Finger, Giuliana Giunta, Alessandro Troisi, and Paola Carbone.
\newblock Framework for a high-throughput screening method to assess polymer/plasticizer miscibility: The case of hydrocarbons in polyolefins.
\newblock {\em Macromolecules}, 57(10):4637--4647, 2024.

\bibitem{Xu2024}
Jiaxin Xu and Tengfei Luo.
\newblock Unlocking enhanced thermal conductivity in polymer blends through active learning.
\newblock {\em npj Computational Materials}, 10(1):74, Apr 2024.

\bibitem{zhang2024active}
Renzheng Zhang, Jiaxin Xu, Hanfeng Zhang, Guoyue Xu, and Tengfei Luo.
\newblock Active learning-guided exploration of thermally conductive polymers under strain.
\newblock {\em Digital Discovery}, 2024.

\bibitem{KIM2021110067}
Chiho Kim, Rohit Batra, Lihua Chen, Huan Tran, and Rampi Ramprasad.
\newblock Polymer design using genetic algorithm and machine learning.
\newblock {\em Computational Materials Science}, 186:110067, 2021.

\bibitem{mannodi2016machine}
Arun Mannodi-Kanakkithodi, Ghanshyam Pilania, Tran~Doan Huan, Turab Lookman, and Rampi Ramprasad.
\newblock Machine learning strategy for accelerated design of polymer dielectrics.
\newblock {\em Scientific reports}, 6(1):20952, 2016.

\bibitem{smith2018less}
Justin~S Smith, Ben Nebgen, Nicholas Lubbers, Olexandr Isayev, and Adrian~E Roitberg.
\newblock Less is more: Sampling chemical space with active learning.
\newblock {\em The Journal of chemical physics}, 148(24), 2018.

\bibitem{zhang2019bayesian}
Yao Zhang et~al.
\newblock Bayesian semi-supervised learning for uncertainty-calibrated prediction of molecular properties and active learning.
\newblock {\em Chemical science}, 10(35):8154--8163, 2019.

\bibitem{RN84}
G.~Giunta, C.~Svaneborg, H.~A. Karimi-Varzaneh, and P.~Carbone.
\newblock Effects of graphite and plasticizers on the structure of highly entangled polyisoprene melts.
\newblock {\em ACS Applied Polymer Materials}, 2(2):317--325, 2020.

\bibitem{RN116}
Giuliana Giunta, Lois Smith, Kristof Bartha, H.~Ali Karimi-Varzaneh, and Paola Carbone.
\newblock Understanding the balance between additives’ miscibility and plasticisation effect in polymer composites: A computational study.
\newblock {\em Soft Matter}, 19(13):2377--2384, 2023.

\bibitem{jorgensen2023opls}
William~L Jorgensen, Mohammad~M Ghahremanpour, Anastasia Saar, and Julian Tirado-Rives.
\newblock Opls/2020 force field for unsaturated hydrocarbons, alcohols, and ethers.
\newblock {\em The Journal of Physical Chemistry B}, 128(1):250--262, 2023.

\bibitem{ghahremanpour2022refinement}
Mohammad~M Ghahremanpour, Julian Tirado-Rives, and William~L Jorgensen.
\newblock Refinement of the optimized potentials for liquid simulations force field for thermodynamics and dynamics of liquid alkanes.
\newblock {\em The Journal of Physical Chemistry B}, 126(31):5896--5907, 2022.

\bibitem{costa2024temperature}
Gabriel~P Costa, Phillip Choi, Stanislav~R Stoyanov, and Qi~Liu.
\newblock The temperature dependence of the hildebrand solubility parameters of selected hydrocarbon polymers and hydrocarbon solvents: a molecular dynamics investigation.
\newblock {\em Journal of Molecular Modeling}, 30(7):196, 2024.

\bibitem{yolong2019local}
Thipjula Yolong and Thana Sutthibutpong.
\newblock Local glass transition of polyisoprene induced by graphene planes observed in silico through atomistic molecular dynamics simulations.
\newblock In {\em Journal of Physics: Conference Series}, volume 1380, page 012081. IOP Publishing, 2019.

\bibitem{chen2022molecular}
Zhiyuan Chen, Qunzhang Tu, Zhonghang Fang, Xinmin Shen, Qin Yin, Xiangpo Zhang, and Ming Pan.
\newblock Molecular dynamics studies of the mechanical behaviors and thermal conductivity of polyisoprene with different degrees of polymerization.
\newblock {\em Polymers}, 14(22):4950, 2022.

\bibitem{RN85}
Henk Bekker, Herman Berendsen, E.~J. Dijkstra, S.~Achterop, Rudi Drunen, David van~der Spoel, A.~Sijbers, H.~Keegstra, B.~Reitsma, and M.~K.~R. Renardus.
\newblock Gromacs: A parallel computer for molecular dynamics simulations.
\newblock {\em Physics Computing}, 92:252--256, 1993.

\bibitem{van2005gromacs}
David Van Der~Spoel, Erik Lindahl, Berk Hess, Gerrit Groenhof, Alan~E Mark, and Herman~JC Berendsen.
\newblock Gromacs: fast, flexible, and free.
\newblock {\em Journal of computational chemistry}, 26(16):1701--1718, 2005.

\bibitem{barton1975solubility}
Allan~FM Barton.
\newblock Solubility parameters.
\newblock {\em Chemical Reviews}, 75(6):731--753, 1975.

\bibitem{RN86}
Lewis~J. Fetters, David~J. Lohse, and William~W. Graessley.
\newblock Chain dimensions and entanglement spacings in dense macromolecular systems.
\newblock {\em Journal of Polymer Science Part B: Polymer Physics}, 37(10):1023--1033, 1999.

\bibitem{svaneborg2016kremer}
Carsten Svaneborg, Hossein~Ali Karimi-Varzaneh, Nils Hojdis, Frank Fleck, and Ralf Everaers.
\newblock Kremer-grest models for universal properties of specific common polymer species.
\newblock {\em arXiv preprint arXiv:1606.05008}, 2016.

\bibitem{signac_commat}
Carl~S. Adorf, Paul~M. Dodd, Vyas Ramasubramani, and Sharon~C. Glotzer.
\newblock Simple data and workflow management with the signac framework.
\newblock {\em Comput. Mater. Sci.}, 146(C):220--229, 2018.

\bibitem{signac_scipy_2018}
Vyas Ramasubramani, Carl~S. Adorf, Paul~M. Dodd, Bradley~D. Dice, and Sharon~C. Glotzer.
\newblock {s}ignac: {A} {P}ython framework for data and workflow management.
\newblock In {\em Proceedings of the 17th Python in Science Conference}, pages 152--159, 2018.

\bibitem{signac_scipy_2021}
Bradley~D. Dice, Brandon~L. Butler, Vyas Ramasubramani, Alyssa Travitz, Michael~M. Henry, Hardik Ojha, Kelly~L. Wang, Carl~S. Adorf, Eric Jankowski, and Sharon~C. Glotzer.
\newblock {s}ignac: {D}ata {M}anagement and {W}orkflows for {C}omputational {R}esearchers.
\newblock In {\em Proceedings of the 20th Python in Science Conference}, pages 23--32, 2021.

\bibitem{10.5555/1593511}
Guido Van~Rossum and Fred~L. Drake.
\newblock {\em Python 3 Reference Manual}.
\newblock CreateSpace, Scotts Valley, CA, 2009.

\bibitem{Lu2023ReBenchmarkingPA}
Po-Yi Lu, Chun-Liang Li, and Hsuan-Tien Lin.
\newblock Re-benchmarking pool-based active learning for binary classification.
\newblock {\em ArXiv}, abs/2306.08954, 2023.

\bibitem{barton2018handbook}
Allan~FM Barton.
\newblock {\em Handbook of Poylmer-Liquid Interaction Parameters and Solubility Parameters}.
\newblock Routledge, 2018.

\bibitem{imre2003effect}
Attila~R Imre, W~Alexander Van~Hook, Bong~Ho Chang, and Young~Chan Bae.
\newblock The effect of alkane chain length on the liquid--liquid critical temperatures of oligostyrene/linear-alkane mixtures.
\newblock {\em Monatshefte f{\"u}r Chemie/Chemical Monthly}, 134:1529--1539, 2003.

\bibitem{yaws2008thermophysical}
Carl~L Yaws.
\newblock {\em Thermophysical properties of chemicals and hydrocarbons}.
\newblock William Andrew, 2008.

\bibitem{lide2004crc}
David~R Lide.
\newblock {\em CRC handbook of chemistry and physics}, volume~85.
\newblock CRC press, 2004.

\end{thebibliography}

\end{document}